\documentclass[3p,citeautoscript]{elsarticle}
\usepackage{lineno,hyperref}
\usepackage{graphicx,rotating,subfigure,amsmath,amsfonts,amssymb,delarray,color}
\usepackage[latin1]{inputenc}
\usepackage{hyperref}
\usepackage{xcolor}
\usepackage{soul}
\usepackage{dsfont}
\usepackage[T1]{fontenc}
\usepackage{physics}
\usepackage{comment}

\modulolinenumbers[5]

\def\12{\frac{1}{2}}

\journal{Annals of Physics}
\biboptions{numbers,sort&compress}

\begin{document}

\begin{frontmatter}

\title{Unlimited growth of particle fluctuations in many-body localized phases}

\author{Maximilian Kiefer-Emmanouilidis}
\address{Department of Physics and Research Center OPTIMAS, University Kaiserslautern, 67663 Kaiserslautern, Germany}
\address{Department of Physics and Astronomy, University of Manitoba, Winnipeg R3T 2N2, Canada}
\author{Razmik Unanyan}
\address{Department of Physics and Research Center OPTIMAS, University Kaiserslautern, 67663 Kaiserslautern, Germany}
\author{Michael Fleischhauer}
\address{Department of Physics and Research Center OPTIMAS, University Kaiserslautern, 67663 Kaiserslautern, Germany}
\author{Jesko Sirker}
\address{Department of Physics and Astronomy, University of Manitoba, Winnipeg R3T 2N2, Canada}
\address{Manitoba Quantum Institute, University of Manitoba, Winnipeg R3T 2N2, Canada}

%
%
%

\begin{abstract}
We study quench dynamics in a t-V chain of spinless fermions
(equivalent to the spin-$1/2$ Heisenberg chain) with strong potential
disorder. For this prototypical model of many-body localization we
have recently argued that---contrary to the established
picture---particles do not become fully localized. Here we summarize
and expand on our previous results for various entanglement measures
such as the number and the Hartley number entropy. We investigate, in
particular, possible alternative interpretations of our numerical
data. We find that none of these alternative interpretations appears
to hold and, in the process, discover further strong evidence for the
absence of localization. Furthermore, we obtain more insights into the
entanglement dynamics and the particle fluctuations by comparing with
non-interacting systems where we derive several strict bounds. We find
that renormalized versions of these bounds also hold in the
interacting case where they provide support for numerically discovered
scaling relations between number and entanglement entropies.
\end{abstract}

\begin{keyword}
Many-Body Localization, Anderson Localization, Disordered Systems, Entanglement Measures
\end{keyword}

\end{frontmatter}


\section{Introduction}

In an Anderson localized (AL) phase of a non-interacting quantum
system, particles are constrained to spatially localized orbitals
\cite%
{Anderson58,AndersonLocalization,AbrahamsAnderson,EdwardsThouless}. As
a consequence, entanglement is short-ranged and there is no
transport. An important question then is what happens if interactions
are added. In the localized eigenbasis of the non-interacting system,
even short-range interactions between the constituent particles will
induce effective non-local interactions and non-local hopping
processes, which could destroy the localized character of the
phase. Numerical investigations of one-dimensional quantum systems
with potential disorder indeed show that interactions can induce a
transition from the AL phase into an interacting ergodic phase
\cite{OganesyanHuse,PalHuse,EnssAndraschkoSirker}. The remaining
conceptual question then is if a localized phase can at all survive
for generic interactions. This question was answered in the
affirmative under certain assumptions using perturbative arguments
\cite{BaskoAleiner,Imbrie2016}. Results from exact diagonalizations
for small Heisenberg chains with magnetic field disorder were also
interpreted as showing a transition from an ergodic phase to a
many-body localized (MBL) phase at some finite critical disorder
strength \cite{OganesyanHuse,PalHuse,Luitz1}. One of the hallmarks of
the putative MBL phase---differentiating it from the AL phase---is
that the entanglement entropy increases logarithmically in time after
a quantum quench from a product state
\cite{ZnidaricProsen,BardarsonPollmann} instead of saturating
quickly. This logarithmic increase finds its explanation in the
effective non-local, exponentially decaying interactions when
transforming the microscopic Hamiltonian into the Anderson basis of
localized orbitals and neglecting long-range hopping processes. The
result is an effective interacting model with exponentially many local
conserved charges \cite{SerbynPapic,HuseNandkishore}. In this
effective model, no hopping processes between the localized orbitals
are present at all. Number fluctuations are therefore bounded and
there is no transport. Assuming that insulating clusters do exist, one
can also construct effective real space renormalization group approaches
to investigate the properties of the ergodic-MBL phase transition
\cite{VoskHusePRX,Goremykina2019,Dumitrescu2019,PotterVasseurPRX,MorningstarHuse}.

This established picture has very recently been challenged on two
fronts: On the one hand, researchers have investigated the properties
of the system near the putative ergodic-MBL phase transition and have
analyzed the scaling of indicators for the transition with system size
and disorder strength
\cite{SuntajsBonca,SuntajsBonca2,SelsPolkovnikov,SelsPolkovnikov2}. 
The results were interpreted as showing that the transition point
shifts to infinite disorder in the thermodynamic limit, with the
conclusion that there is no MBL phase for finite disorder. On the
other hand, particle fluctuations deep in the putative MBL phase have
been studied by us using measures such as the number entropy $S_N$ and
the Hartley number entropy $S_H$
\cite{KieferUnanyan1,KieferUnanyan2,KieferUnanyan3}. For all disorder
strengths and system sizes we were able to study numerically in these
papers, we have found that the number entropy does not saturate as
expected based on the established picture for MBL phases, but rather
continues to increase as $S_N\sim\ln\ln t$ after the quantum
quench. We found this to be true not only for the average but also for
the median $S_N$. The observed double logarithmic scaling in time
therefore appears to be the typical behavior and not related to rare
configurations. This points to an absence of true localization due to
a continuing, albeit subdiffusive, transport of particles. The system
appears to remain ultimately ergodic. Our philosophy here is to study
the behavior deep in the putative MBL phase rather than close to the
putative phase transition where finite-size effects are expected to be
most severe. We will briefly discuss possible scenarios for the complete phase
diagram of the model in the conclusions.

One potential problem with the above mentioned results is that they
are all based on numerical data for relatively small system
sizes. While much larger system sizes and even systems in the
thermodynamic limit can be investigated using matrix product states
(MPS)
\cite{ZnidaricProsen,AndraschkoEnssSirker,EnssAndraschkoSirker,Doggen2018,Doggen2019}, 
the build-up of entanglement makes it then impossible to investigate
quench dynamics at long times. It is therefore important to carefully
study the scaling with system size and disorder strength. Trying to
investigate the stability of the MBL phase based on a scaling at or
near the critical point, however, might be particularly prone to
finite size issues, and it cannot be excluded that the behavior in the
thermodynamic limit can only be inferred from studying much larger
systems. Such an argument based on a comparison with models where
analytical results are available or where larger system sizes were
explored have recently been made in
Refs.~\cite{Abaninrecent,Sierant2020,Buijsman2020} with regard to the
results by Suntajs \textit{et al} \cite{SuntajsBonca}. Our results for
the number and Hartley entropies in
Refs.~\cite{KieferUnanyan2,KieferUnanyan3}, on the other hand, were
obtained for disorder strengths which supposedly are deep in the MBL
phase, where finite size issues should be much less severe. There are
nevertheless also at least four possible issues with the
interpretation of our results: (1) The observed increase of the
particle fluctuations might be transient. I.e., the expected
saturation only sets in at longer times. (2) The critical disorder
strength is much larger and the MBL phase thus much smaller than
anticipated. This could mean that the observed behavior is indicative
of the phase transition and not of the MBL phase. (3) The dynamics of
the system in the MBL phase but still relatively close to the
transition might be prone to effects of rare disorder configurations
and our observations might be a result of those. (4) Particle
fluctuations could potentially be very slow in building up but
strictly limited in space. In other words, the observed slow increase
of $S_N(t)$ could be a result of very few particles near the boundary
fluctuating back and forth between the two subsystems. Criticism along
some of these lines has been put forward in Ref.~\cite{LuitzBarLev}.

The main objectives of this article are to (a) summarize and expand on
the results presented by us in
Refs.~\cite{KieferUnanyan1,KieferUnanyan2,KieferUnanyan3}, and (b) to
address the possible issues mentioned above. Our paper is organized as
follows: In Sec.~\ref{Sec_Model}, we introduce the model that we
investigate, define the entanglement measures, and describe the
numerical methods used. We then present in Sec.~\ref{Sec_fluc} our
results for the number entropy $S_N(t)$ and Hartley entropy
$S_H(t)$. We analyze the scaling with system size and disorder
strength and address the influence of rare configurations. One can
gain further insights by comparing the results to free disordered
systems where exact bounds for the entanglement measures can be
derived. This will be done in Sec.~\ref{Sec_free}. The last section is
devoted to a summary and conclusions.


\section{Model, Entropies, and Methods}

\label{Sec_Model} 

We will concentrate on investigating the one-dimensional t-V model 
\begin{equation}  \label{Ham}
H = -J\sum_j(c_j^\dagger c_{j+1} +h.c.) + V \sum_j n_j n_{j+1} + \sum_j D_j
n_j \, .
\end{equation}
Here $J$ is the nearest-neighbor hopping amplitude (we reserve $t$ for
time), $V$ the nearest-neighbor interaction, and $D_j\in [-D/2,D/2]$ a
random onsite potential describing diagonal disorder. We set $\hbar=1$ and
assume $J=1$ thus fixing our time unit as $J^{-1}$. $n_j=c_j^\dagger c_j$ is
the particle number at site $j$. Note that this model is equivalent to a
spin-$1/2$ XXZ Heisenberg chain with magnetic field disorder. We will
concentrate on $V=2J$ which corresponds to the isotropic Heisenberg model.
The investigated chains have open boundary conditions and an even number of
sites at half filling. The system is prepared in an initial product state $|\Psi(0)\rangle$
and we numerically calculate the time evolved state $|\Psi(t)\rangle
=\exp(-iHt)|\Psi(0)\rangle$. From this we determine the reduced density
matrix $\rho_A$ by splitting the system into two equal halves, $A$ and $B$,
and tracing out one subsystem, $\rho_A(t) = \tr_B |\Psi(t)\rangle\langle
\Psi(t)|$. 

Our main measures to investigate the ensuing quench dynamics are
entanglement and number entropies. We define, in particular, the R\'enyi
entropy of order $\alpha$ as 
\begin{equation}  \label{Renyi}
S^{(\alpha)} = \frac{\ln\tr\rho_A^\alpha}{1-\alpha} \, .
\end{equation}
The von-Neumann entanglement entropy is obtained by $S=\lim_{\alpha\to 1}
S^{(\alpha)}$. Since the total particle number is conserved, there are two
distinct sources for entanglement. One is due to superpositions of different
configurations for a fixed particle number $n$ in subsystem $A$. This part
is the configurational entropy $S^{(\alpha)}_{\text{conf}}$. The other
source of entanglement are particle number fluctuations between the two
subsystems. To characterize this type of entanglement we define the R\'enyi 
\emph{number} entropy 
\begin{equation}  \label{NumberEnt}
S_N^{(\alpha)} = \frac{\ln \sum_n \Bigl(p(n)\Bigr)^\alpha}{1-\alpha}
\end{equation}
where $p(n)$ is the probability of finding $n$ particles in subsystem $A$.
The total R\'enyi entanglement entropy is the sum of the two contributions, $%
S^{(\alpha)} = S^{(\alpha)}_N + S^{(\alpha)}_{\text{conf}}$. For $\alpha\to
1 $ one obtains, in particular, the following splitting of the von-Neumann
entanglement entropy \cite%
{KlichLevitov,WisemanVaccaro,DowlingDohertyWiseman,SchuchVerstraeteCirac,SchuchVerstraeteCirac2,SongFlindt,Rakovszky2019,parez2020,SongRachel,Bonsignori2019,MurcianodiGiulio,MurcianodiGiulio2,LukinRispoli}
\begin{equation}  \label{SvN}
S = S_N + S_{\text{conf}} = -\sum_n p(n) \ln p(n) -\sum_n p(n)\tr%
\{\rho_A(n)\ln\rho_A(n)\}
\end{equation}
where $\rho_A(n)$ is the block of the reduced density matrix with particle
number $n$.

We note that the entanglement measures defined here are not only useful to
study quench dynamics numerically but are also accessible in experiments on
cold atomic gases and trapped ions. Number entropies, in particular, can be
easily accessed for any experimental system where particle number
spectroscopy with single site resolution is possible. Obtaining the particle
number distribution function $p(n)$ at time $t$ after the quench then simply
amounts to counting the number of particles in subsystem $A$ at this time
and repeating the experiment many times to obtain a good statistics.
Measuring either $S^{(\alpha)}$ or $S_{\text{conf}}^{(\alpha)}$
experimentally --- once the corresponding number entropy is known both
quantities give the same information --- is typically much harder. One
possibility is a full quantum tomography but this method is very time
consuming and limited to small system sizes. Very recently, two alternatives
have emerged: On the one hand, it was shown in Ref.~\cite{LukinRispoli} that 
$S_{\text{conf}}$ can be approximated in a system with weak overall
entanglement by a configurational correlator. On the other hand, it was also
shown recently that the second R\'enyi entropy $S^{(2)}$ can be measured in
a trapped ion system in a way which is more efficient than a full quantum
tomography \cite{BrydgesElben}.

In the following, we will evaluate the entanglement measures above for
the t-V model based on exact diagonalizations (ED) and a
Trotter-Suzuki decomposition of the time evolution operator
\cite{Trotter,Suzuki1,Suzuki2}.  In the former case we treat chains up
to lengths $L=14$ while we can consider chains up to $L=24$ in the
latter case. The study of even longer chains is in principle possible,
however, the need to calculate thousands of disorder samples to obtain
disorder averaged quantities then results in a prohibitive amount of
required computing time even on supercomputers with the latest
generation of graphical processing units (GPUs). While for ED the time
up to which reliable results can be obtained is only limited by the
numerical precision---here we use double precision limiting times to
$t\lesssim 10^{14}$ in units of the inverse hopping amplitude
$J^{-1}$---the Trotter-Suzuki decomposition leads to a decomposition
error which accumulates over time. For the Trotter-Suzuki parameters
$\varepsilon$ chosen here---with $t=\varepsilon N$ and $N$ being the
Trotter-Suzuki number---we are limited to times $t\lesssim 10^4$. We
will specify the system sizes used, the number of realizations, and
the initial states in the captions of the corresponding figures. If
not otherwise specified, we average all quantities by computing the
measure for each realization first and then average over all
realizations, e.g.~$S_N = -\overline{\sum_n p(n)\ln p(n)}$.


\section{Slow growth of particle fluctuations in MBL phases}

\label{Sec_fluc} 



\begin{figure}[tbp]
\includegraphics[width=1\textwidth]{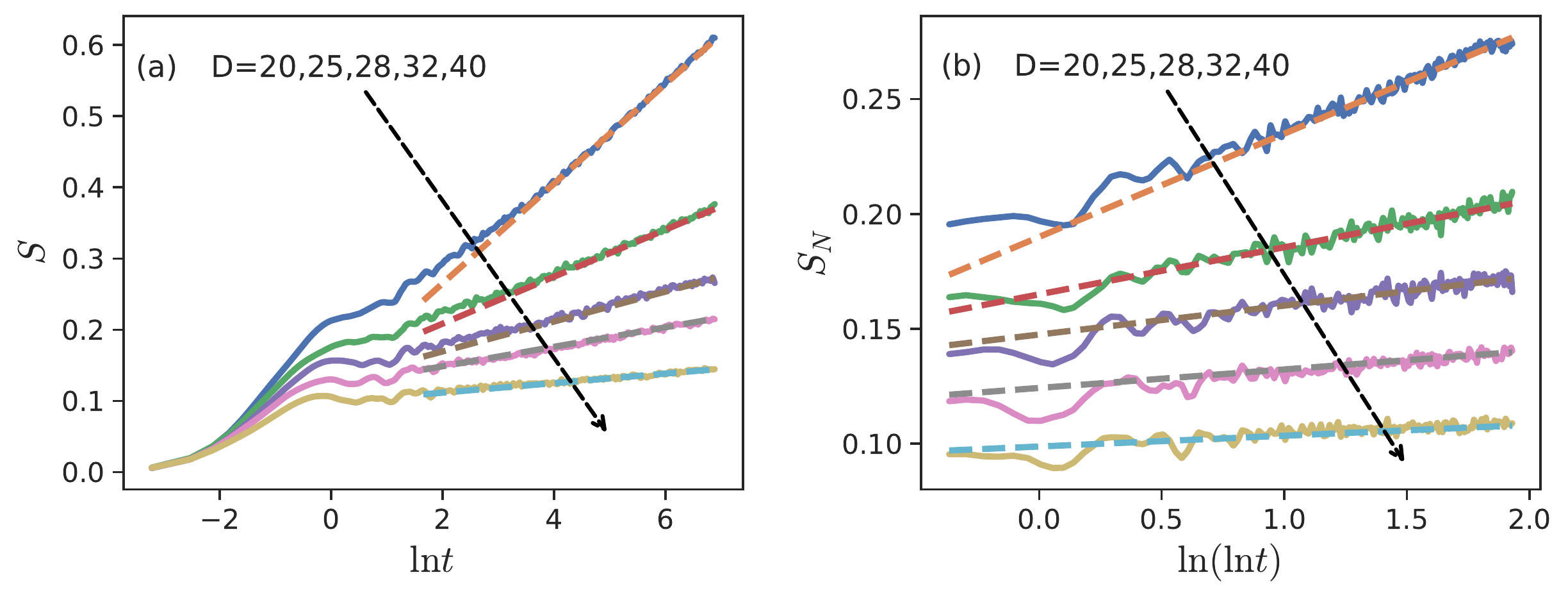}
\caption{Numerical results for (a) the von-Neumann entanglement entropy, and
(b) the number entropy for strong disorder, $D>D_c\sim 16$, for a
system with $L=24$ sites. The dashed lines represent logarithmic fits
in (a) and double logarithmic fits in (b). Here we averaged over 1500
disorder realization for $D\leq 28$ and 2000 for $D> 28$ starting from
a random half-filled product state. Panel (b) is based on data already
presented in Fig.~3 of Ref.~\cite{KieferUnanyan3}.}
\label{Fig1}
\end{figure}

We will start by reviewing our main result as obtained recently in
Refs.~\cite{KieferUnanyan2,KieferUnanyan3}. According to previous
numerical studies \cite{OganesyanHuse,PalHuse,Luitz1,Luitz2}, the
model \eqref{Ham} shows a phase transition from an ergodic to a
putative MBL phase at a critical disorder strength $D_c\sim 16$. In
Fig.~\ref{Fig1}, we show results for the total von-Neumann entropy and
the number entropy for disorder strengths $D>D_c$.
%
%
For the von-Neumann entanglement entropy we find $S(t)\sim \ln t$
consistent with previous studies
\cite{ZnidaricProsen,BardarsonPollmann,AndraschkoEnssSirker}. Surprisingly,
however, the number entropy also seems to grow without bounds and is
well described by $S_N(t)\sim\ln\ln t$. This apparently contradicts
the very notion of a localized phase: The number entropy has to
saturate if the motion of particles is limited to a finite region in
space. To be more precise, the number entropy is a measure describing
how broad the particle number distribution $p(n)$ in subsystem $A$
is. Here we consider an initial product state with $L/2$ particles of
which $n_{\textrm{ini}}$ particles are initially in subsystem $A$. The
initial number entropy is therefore zero. The maximal number entropy
is obtained if each possible number of particles $n=0,1,\cdots,L/2$ in
subsystem $A$ has the same probability leading to
$S_N^{\text{max}}=\ln(L/2+1)$. If the particles are localized,
however, then only those particles originally situated close to the
boundary should be able to cross from one subsystem to the other. If,
for example, only fluctuations with
$n=n_{\textrm{ini}},n_{\textrm{ini}}\pm 1$ are possible, then the
number entropy would be bounded by $S_N\leq\ln 3$.

Let us now address the points of possible criticism mentioned in the
introduction.


\subsection{Closeness to criticality}
\label{Close}

\begin{figure}[tbp]
\begin{center}
\includegraphics[width=1\textwidth]{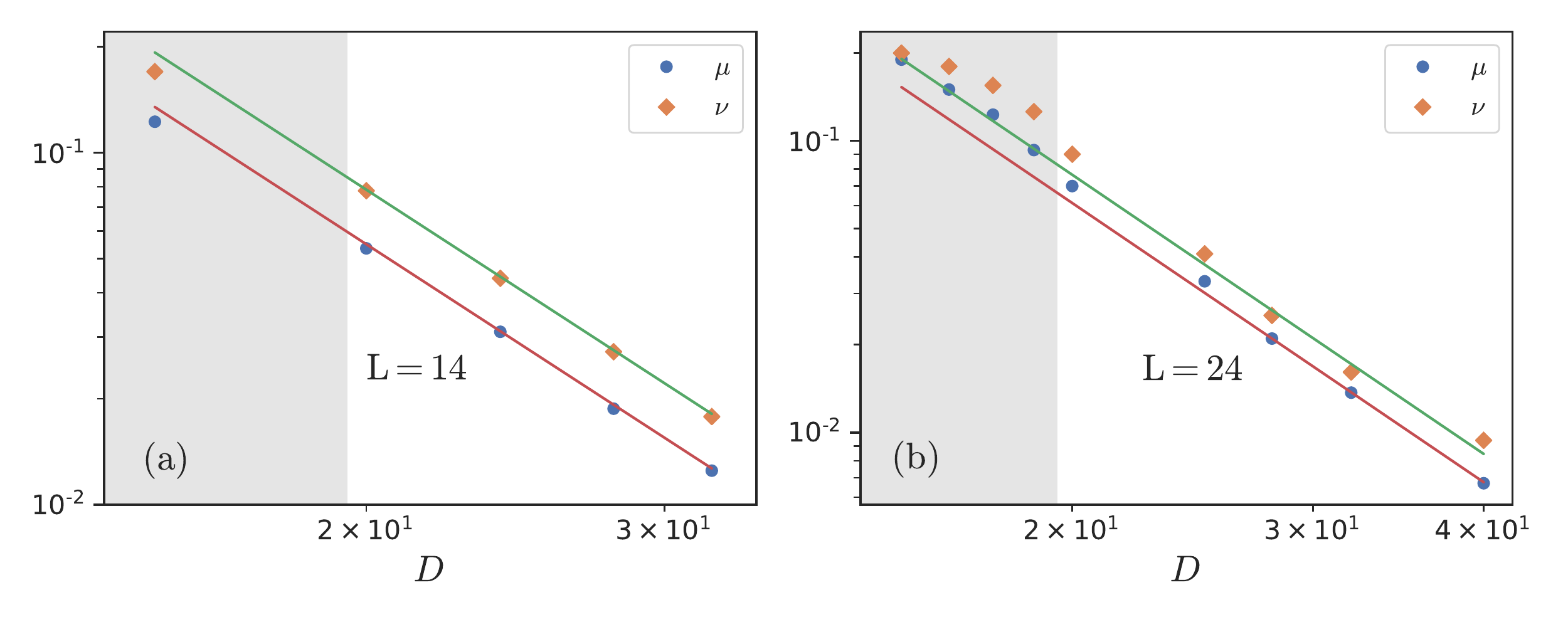}
\end{center}
\caption{The data in Fig.~\protect\ref{Fig1} are fitted by $S=\protect\mu\ln
t$ and by $S_N=\frac{\protect\nu}{2}\ln\ln t$. The plots show the
prefactors $\protect\mu,\protect\nu$ extracted from these fits for a
system size of (a) $L=14$ and (b) $L=24$ as a function of disorder
strength $D$. The lines correspond to power-law fits $\sim
1/D^{\protect\alpha}$ of $\mu,\nu$ deep in the MBL regime with
$\protect\alpha
\approx 3$. The shaded area indicates the disorder regime where the
system might be critical and the values of $\protect\mu ,\protect\nu$
in this regime should be considered with care. For $L=14$ we simulated
10,000 disorder realizations starting from 50 random half-filled
product states and in the case of $L=24$ we simulate 1500 disorder
realization for $D\leq 28$ and 2000 for $D > 28$ starting from a
single random half-filled product state.}
\label{Fig2}
\end{figure}
\begin{figure}[tbp]
\includegraphics[width=1\textwidth]{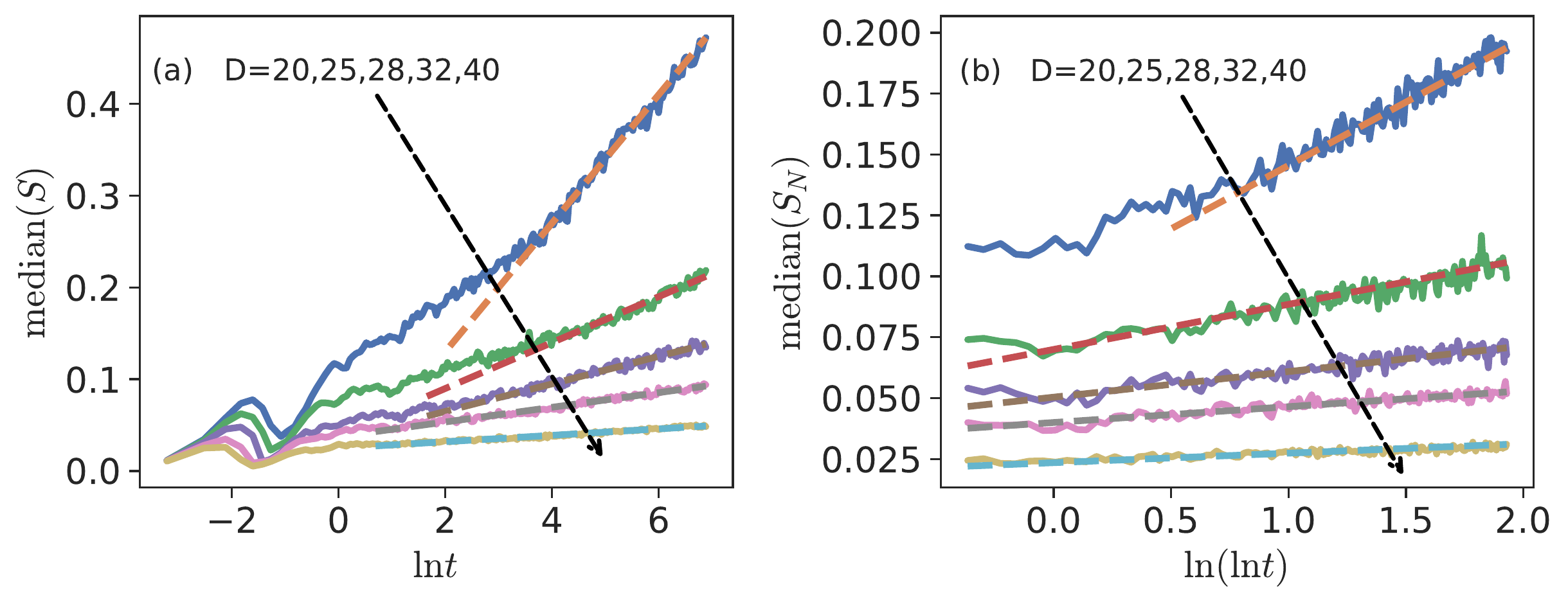}
\caption{Numerical results for (a) the median of the von-Neumann entanglement entropy, and
(b) the median of the number entropy for strong disorder, $D>D_c\sim 16$,
for a system with $L=24$ sites. The dashed lines represent logarithmic
fits in (a) and double logarithmic fits in (b). Here we calculate the
median from 1500 disorder realization for $D\leq 28$ and from 2000
realizations for $D> 28$ starting from a random half-filled
product state.}
\label{Fig1a}
\end{figure}

The first possible issue might be that the observed $S_N\sim\ln\ln t$
is a consequence of being too close to the ergodic-MBL transition
where rare configurations might strongly influence the dynamics
\cite{Gopalakrishnan2015,Agarwal2017}. Here we note first that this
behavior is observed for disorder strengths ranging from values close
to the transition all the way up to $D=40$, which is $\sim 2.5D_c$,
see Fig.~\ref{Fig1}(b). The only change with increasing disorder
strength is that the prefactor of the $\ln\ln t$ scaling becomes
smaller as can be seen in Fig.~\ref{Fig2}.
%
%
Most importantly, the time dependence of neither $S$ nor $S_N$ changes
qualitatively. It is known that within the MBL phase but close to the
transition rare regions with less disorder can cause a very slow dynamics
\cite{Gopalakrishnan2015,Agarwal2017} and can destabilize the MBL
phase in small systems. In order to exclude such a scenario, we also
calculated the median of the entanglement and number entropies, shown
in Fig.~\ref{Fig1a}. The median quantities are defined by sorting the
entropies for each realization in terms of magnitude at every point in
time and then choosing the value in the middle, for an odd number of
realizations, or the average of the two middle values, for an even
number of realizations. The median number entropy shows the same
double logarithmic growth in time as the average number entropy, shown
in Fig.~\ref{Fig1}(b). We conclude that the observed long-time growth
is not the consequence of rare regions but rather represents the
typical behavior of the number entropy. The main qualitative
difference between averaged and median entropies is a suppression of
the initial increase in the median as compared to the average,
i.e.~rare regions do influence the short-time behavior but not the
long-time scaling. More details about the dependence of $S_N$ on the
disorder realizations are discussed in \ref{AppA}.

The data thus do not support the notion that the growth of $S_N(t)$
changes in a qualitative manner if we move deeper into the putative
MBL phase. The observed $S_N\sim\ln \ln t$ scaling rather seems to be
an intrinsic property of the MBL phase---at least for the simulation
times we are able to achieve numerically. Fig.~\ref{Fig2} also
indicates that the scaling of the total entanglement entropy $S$ and
that of the number entropy $S_N$ are very closely linked. The
prefactors of the logarithmic and double logarithmic fits show
\textit{the same} power-law dependence on disorder.

%
%


\subsection{Scaling with system size and disorder strength}

\begin{figure}[tbp]
\begin{center}
\includegraphics[width=1\textwidth]{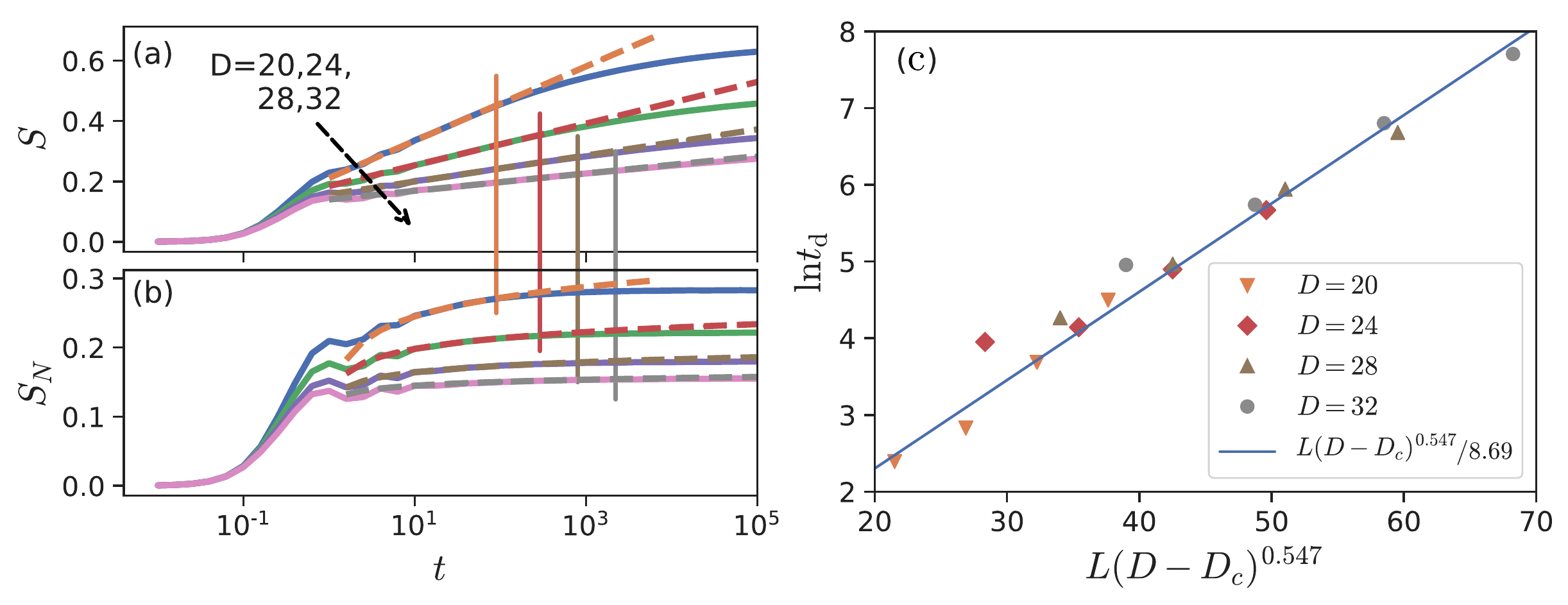}
\end{center}
\caption{The time $t_d$ for $L=14$ where the numerical data for $S$ start to
deviate from the fit $S=\protect\mu\ln t$, shown in (a), is the same
time where $S_\mathrm{N}$ starts to deviate from the fit
$S_N=\frac{\protect\nu}{2}\ln\ln t $, shown in (b). The data has been
adopted from Fig.~2 in Ref.~\cite{KieferUnanyan3}. (c) $t_d$ depends
on $L/\ell$ only with $\ell=(D-D_c)^{-\protect\alpha}$, $D_c\approx
16$, and $\protect\alpha\approx 0.5$ leading to a scaling
collapse. Over 10,000 disorder realizations starting from 50 random
initial half-filled product states have been simulated. For the fitting process in (c), data for system sizes $L=10,12,14$ have been used.}
\label{Fig3}
\end{figure}

Another question one might raise with regard to the interpretation of
the results shown in Fig.~\ref{Fig1} and Fig.~\ref{Fig1a}, is whether
the increase of $S_N$ is transient and will ultimately give way to
saturation. A problem in addressing this issue is, of course, the
limited system sizes which are amenable to a numerical solution. The
best evidence that the behavior is not transient, is based on the
following observation: For every system size $L$ and any $D>D_c$ there
is a time $t_d$ where the numerical data start to deviate from
$S_N=\frac{\nu}{2}\ln\ln t$ due to finite size effects. This is
illustrated in Fig.~\ref{Fig3}(a,b) which also shows that this time $t_d$
is the same time where also the data for $S$ start to deviate from a
logarithmic scaling, further supporting the notion that the growth of
$S_N$ is linked to the growth of $S$. We have extracted the time $t_d$
from the numerical data and find that $t_d\sim
\exp(L/\ell)$ where $\ell\sim(D-D_c)^{-\alpha}$ is a characteristic
length scale. This relation is illustrated in Fig.~\ref{Fig3}(c)
where we observe an almost perfect scaling collapse of $t_d$ as
function of $L/\ell$.
We believe this to be a very strong indication that the observed growth of
the number entropy is not transient. Based on this analysis, we expect that
in the thermodynamic limit $S_N$ grows without bounds throughout the
putative MBL phase.


\subsection{$p(n)$ and truncated Hartley number entropy}


Next, we want to address the possible criticism that the time regime
in which we observe the double logarithmic scaling of the number
entropy is a regime where $S_N<\ln 3$. The increase of the number
entropy thus could potentially be explained by a single particle
fluctuating between the two subsystems. In order to investigate this
point, we have to consider the full particle number distribution
$p(n)$. If indeed only small fluctuations around the initial particle
number $n_\text{ini}$ in the subsystem contribute, then we expect that
the distribution only changes in time for particle numbers close to
this initial value, while $p(n)$ remains exponentially small for large
fluctuations of $n$ away from this value at all times.

\begin{figure}[tbp]
\begin{center}
\includegraphics[width=1\textwidth]{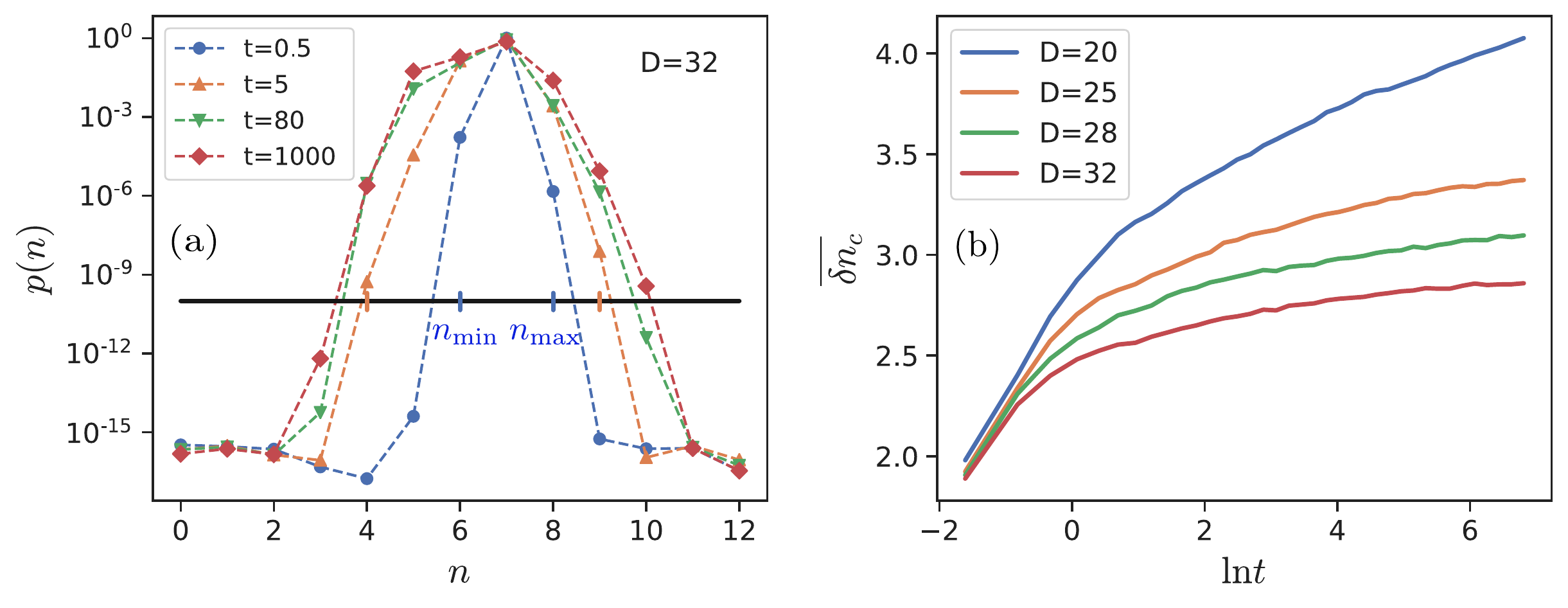}
\end{center}
\caption{(a) Definition of the width $\protect\delta n_c$ shown for a
particle number distribution of a single disorder realization for
$D=32,\, L=24$, and a cutoff $p_c=10^{-10}$. The ticks on the cutoff
line show exemplarily $n_\mathrm{min}$ and $n_\mathrm{max}$ for $t=0.5$
and $t=5$. (b) Time evolution of the average width $\overline{\delta n_c}$ for a cutoff
$p_c=10^{-10}$ for various disorder strengths. Here we simulated 1500
disorder realization for $D\leq 28$ and 2000 for $D > 28$ starting
from a random half-filled product state.}
\label{Fig4}
\end{figure}

To investigate the change of the particle number distribution in time,
we define for each sample its width by $\delta n_c =
n_{\text{max}}-n_{\text{min}}$ with $p(n_{\text{max/min}})>p_c$. This
is shown for one particular sample in Fig.~\ref{Fig4}(a).
While the width $\delta n_c$ does depend on the value chosen for $p_c$, we
find that the scaling of the average width is always given by $\overline{\delta n_c} \sim
(\ln t)^\nu$ with some positive exponent $\nu$ provided that $p_c\ll 1$.
This growth of the width of the particle number distribution is shown for
different disorder strengths in Fig.~\ref{Fig4}(b). It is a clear indication
that large particle number fluctuations do occur and that changes of $p(n)$
in time \textit{are not} limited to redistributions close to $n=n_{\text{ini}%
}$ as would be expected if the MBL phase is truly localized.

\begin{figure}[tbp]
\includegraphics[width=1\textwidth]{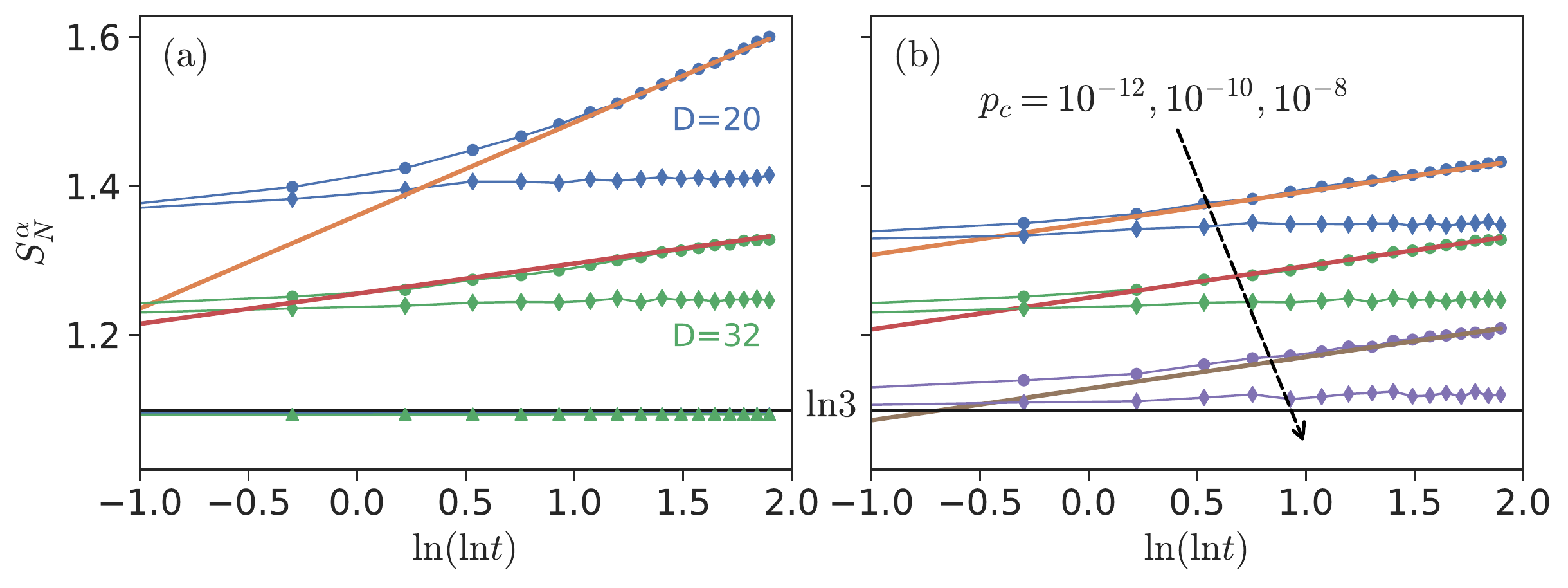}
\caption{Truncated Hartley number entropies: (a) $S_N^{(\protect\alpha)}$ 
with $\protect\alpha=10^{-3}$ and $p_c=10^{-10}$ for the MBL case
(circles) and the Anderson case (diamonds) for $L=24$. The lines are
double logarithmic fits. Also shown are the entropies when only
$p(\bar n)$ and $p(\bar n\pm 1)$ are taken into
account (triangles). The latter saturate at $\ln 3$. (b) Dependence of
the results on the cutoff $p_c$ for a disorder strength $D=32$. Both
figures are based on data already used in Fig.~5 and Fig.~6 in
Ref.~\cite{KieferUnanyan3}. 1500 disorder realizations for $D\leq 28$
and 2000 for $D > 28$ have been used starting from a single random
half-filled product state.}
\label{Fig5}
\end{figure}

Another way to see this, is to study the Hartley number entropy $S_H =
\lim_{\alpha\to 0} S_N^{(\alpha)}$ \cite{KieferUnanyan3}. The Hartley number
entropy counts the particle numbers $n$ for which $p(n)\neq 0$.  Since
a unitary time evolution will immediately lead to a non-zero
probability for any particle distribution consistent with the
conservation laws, independent of whether or not the system is
localized, it is important to introduce a cutoff $p_c>0$ and to only
consider configurations with $p(n,t)>p_c$. All values below the cutoff
are set to zero and the distribution is renormalized. If the system is
in a localized phase, this truncated Hartley number entropy for {\it
any} cutoff $p_c>0$ has to saturate in the thermodynamic limit at a
value which is much below the equipartition value, corresponding to a
fully thermalized infinite-temperature state. Only in the limit $p_c
\to 0$ will $S_H$ asymptotically approach the equipartition value and a
discrimination from an ergodic phase is no longer possible. Thus it is
important to consider a \emph{truncated} Hartley entropy with a
non-zero cutoff $p_c$. We here choose a threshold $p_c$ which is well
above the accuracy of our numerical calculations, which are done in
double precision. Note that a relatively large cutoff will suppress
the Hartley number entropy and make a distinction from the
Anderson case impossible. Furthermore, we cannot take the limit
$\alpha\to 0$ exactly numerically but rather consider a small but
finite value of $\alpha=10^{-3}$. Results for the strongly disordered
t-V model \eqref{Ham} with $V=2$ (MBL case) are compared to $V=0$
(Anderson case) in Fig.~\ref{Fig5}.
There is a clear qualitative difference. While the Hartley number
entropy quickly saturates in the Anderson case---showing that the
possible particle numbers $n$ in the subsystem with $p(n)>p_c$ are
limited, consistent with localization---$S_H$ continues to grow
$\sim\ln\ln t$ similar to the number entropy shown in Fig.~\ref{Fig1}
and Fig.~\ref{Fig1a}. In addition, we also show in Fig.~\ref{Fig5} the
entropies if only configurations with $p(\bar{n})$ and $p(\bar n\pm
1)$ are taken into account where $\bar n$ is the particle number with
maximal probability. It is obvious, that these configurations alone
cannot explain the observed growth of $S_H$. Finally, we note that
while the cutoff $p_c$ does quantitatively change the results it does
not change the $S_H\sim\ln\ln t$ growth.

In conclusion, the data presented in Fig.~\ref{Fig4} and
Fig.~\ref{Fig5} clearly show that the observed increase of the
particle number fluctuations after the quench cannot be explained by a
small number of particles fluctuating between the subsystems. Instead,
the probability for large particle number fluctuations is continuously
growing in time in the putative MBL phase. This is inconsistent with a
true localization of particles and is very different from the behavior
observed in the Anderson localized phase. As a next step, we will try
to shed some further light on the link between the von-Neumann entropy
and the number entropy by deriving exact bounds for non-interacting
fermionic and bosonic systems.


\section{Bounds for number entropies and relation to particle fluctuations}

\label{Sec_free} 
From Fig.~\ref{Fig1} , Fig.~\ref{Fig1a}, and Fig.~\ref{Fig3} we have
seen that the $S\sim \ln t$ and $S_{N}\sim \ln \ln t$ growths of
von-Neumann and number entropies seem to be linked. Due to the
limiting procedure involved in obtaining $S$ and $S_N$, these quantities
are difficult to work with in analytical calculations. Instead, we
will concentrate on the second Rényi number entropy $S_{N}^{\left(
2\right) }$ and the second Rényi entropy $S^{\left( 2\right) }$ and
start by considering Gaussian fermionic and bosonic systems. For these
systems, we derive bounds for $S_N^{\left( 2\right) }$ in terms of
$S^{\left( 2\right) }$.  We will also clarify the connection between
the number entropy and particle fluctuations $\Delta N=\sqrt{\langle
N^{2}\rangle -\langle N\rangle ^{2}}$. After deriving these exact
relations for non-interacting systems, we will return to the
interacting t-V model and show that similar relations also appear to
hold in this case.


\subsection{Exact bounds for free fermions and free bosons}

In order to derive upper and lower bounds on the second R\'enyi number
entropy $S_{N}^{\left( 2\right) }$ in terms of particle fluctuations
and the second R\'enyi entropy $S^{\left( 2\right) }$, we make use of
the fact that the quantum state for a non-interacting fermionic or
bosonic system in any dimension is completely determined by its
single-particle correlations and has a Gaussian form
\cite{Peschel2004,PeschelEisler}. Since we assume, furthermore, total
particle number conservation, the density matrix $\rho$ can be
represented as
\begin{equation}
\rho=\frac{1}{\mathcal{Z}}\mathtt{tr}\left[ \exp\left( -{\displaystyle%
\sum\limits_{m,n}} c_{m}^{\dagger}C_{mn}c_{n}\right) \right] ,
\label{density_matrix}
\end{equation}
where $c_{m}\left( c_{m}^{\dagger}\right) $ are the fermionic or bosonic
annihilation (creation) operators at lattice site $m$. Here $\mathbf{C}$ is a
Hermitian matrix which is determined entirely by single-particle
correlations. The partition function $\mathcal{Z}=$ $\mathtt{tr}\Bigl[ \exp\Bigl(
-{\displaystyle\sum\limits_{m,n}} c_{m}^{\dagger}C_{mn}c_{n}\Bigr)
\Bigr] $ then reads 
\begin{equation}
\mathcal{Z}=\det\left( 1-se^{-\mathbf{C}}\right) ^{-s},
\label{partion_function}
\end{equation}
where $s=1$ for bosons and $s=-1$ for fermions.

It is useful to introduce the moment generating function \cite%
{KlichLevitov09} of the total particle number $N=\sum\limits_{m}
c_{m}^{\dagger}c_{m}$ in the considered partition in the form
\begin{equation}
\chi\left( \theta\right) =\mathtt{tr}\Bigl( \rho\exp\left( -i\theta N\right) %
\Bigr) ={\displaystyle\sum\limits_{n=0}^{\infty}} p(n)\exp\left( -i\theta
n\right) ,  \label{characteristic_function}
\end{equation}
whose Fourier coefficients are the probabilities $p(n)$ to find $n$
particles in the subsystem. It encodes all the information about the
particle statistics. The moments of the distribution are given by the
coefficients of the logarithm $\chi\left( \theta\right) $ \cite%
{KlichLevitov09,SongFlindt,CalabreseMintchev}. In particular, the
average particle number in the subsystem is given by
\begin{equation}
\langle N\rangle = {\displaystyle\sum\limits_{n=0}^{\infty}} p(n)\, n=i\frac{%
\partial}{\partial\theta}\ln\chi\left( \theta\right) \bigr\vert_{\theta=0} \,
\label{average}
\end{equation}
and the particle-number variance by 
\begin{equation}
\Delta N^{2}= {\displaystyle\sum\limits_{n=0}^{\infty}} p(n)\bigl( n-\langle
N\rangle\bigr)^{2}=\left( i\frac{\partial}{\partial\theta}\right)
^{2}\ln\chi\left( \theta\right) \bigr\vert_{\theta=0} \, .
\label{fluctuations}
\end{equation}
%

\subsubsection{Free fermions}

For Gaussian fermionic states, the generating function can be written as a
determinant 
\begin{equation}
\chi\left( \theta\right) =\det\left[ \mathbf{1}+\left( e^{-i\theta }-1\right) 
\frac{\mathbf{1}}{\mathbf{1}+e^{\mathbf{C}}}\right] .
\label{characteristic_Deter}
\end{equation}
Making use of Parseval's theorem one then finds 
\begin{equation}
\sum_{n=0}^{\infty} p(n)^{2}=\frac{1}{2\pi} \int_{-\pi}^{\pi}d\theta\,
\left\vert \chi\left( \theta\right) \right\vert ^{2}= \frac{1}{2\pi}
\int_{-\pi}^{\pi} d\theta\det\Bigl( \mathbf{1}-\mathbf{G}\left(
1-\cos\theta\right) \Bigr),  \label{purity}
\end{equation}
where 
\begin{equation}
\mathbf{G=}\frac{2e^{\mathbf{C}}}{\left( \mathbf{1}+e^{\mathbf{C}}\right)
^{2}}  \label{correlation}
\end{equation}
is a positive definite matrix and all eigenvalues are bounded by $1/2$. It
is remarkable that, according to Eq.(\ref{fluctuations}), \ the number
fluctuation $\Delta N^{2}$ are related to the eigenvalues of $\mathbf{G}$ in
the following simple way 
\begin{equation}
\Delta N^{2}=\frac{\mathtt{tr}\,\mathbf{G}}{2}.  \label{Fluctuation_G}
\end{equation}
In order to derive bounds for $S_{N}^{\left( 2\right) }$ in terms of $%
S^{\left( 2\right) }$ we also need a relationship between the correlation
matrix $\mathbf{G}$ and the second R\'enyi entropy $S^{\left( 2\right) }$ 
\begin{equation}
S^{\left( 2\right) }=-\ln\mathtt{tr}\rho^{2}=-\mathtt{tr}\ln\left( 1-\mathbf{%
G}\right) .  \label{Renyi_Second_Renyi}
\end{equation}
Now we are ready to derive the promised bounds for $S_{N}^{\left( 2\right) }$
in terms of the number fluctuations and the second R\'enyi entropy.

\paragraph{Upper bound}

By using the identity $\det A=\exp(\mathtt{tr}\ln A)$ for any positive
definite matrix $A$, expression (\ref{purity}) can be written as 
\begin{eqnarray}
\sum_{n=0}^{\infty} p(n)^{2} &=& \frac{2}{\pi} \int_{0}^{\pi/2}\!\!\!
d\theta\, \exp\left[ \mathtt{tr}\, \ln\left( \mathbf{1}-2\mathbf{G}%
\sin^{2}\theta\right) \right] \\
&\geq& \frac{2}{\pi}\int_{0}^{\pi/2}\!\! \! d\theta\, \exp\left[ 2\mathtt{tr}%
\, \mathbf{G}\ln\left( 1-\sin^{2}\theta\right) \right] =\frac{\Gamma\left( 
\frac{1}{2}+4 \Delta N^{2}\right) }{\sqrt{\pi }\Gamma\left( 1+4 \Delta
N^{2}\right) },  \label{Purity_Gamma}
\end{eqnarray}
where in the last line we have used the fact that $2\mathbf{G}\leq1$ and the
inequality 
\begin{equation}
x \frac{\ln (1-b)}{b} \geq \ln\left( 1-x\right) \geq x\frac{\ln\left( 1-a\right) }{a},
\label{log_Inequality}
\end{equation}
which holds for $0\leq b\leq x\leq a\leq1$. These inequalities follow
simply from the fact that the function $\frac{\ln (1-x)}{x}$ is a
monotonously decreasing function of $x$. Then the integral can be
calculated elementary in terms of the gamma function,
i.e. $\frac{2}{\pi}%
\int_{0}^{\pi/2} d\theta\, \cos^{p}\theta =\frac{\Gamma\left( \frac{1+p}{2}%
\right) }{\sqrt{\pi }\Gamma\left( 1+\frac{p}{2}\right) },$ for $p>-1$. 
The inequality (\ref{Purity_Gamma}) yields the following bound
\begin{equation}
S_{N}^{\left( 2\right) }=-\ln\left( \sum_{n} p(n)^{2}\right) \leq\ln\frac{%
\sqrt{\pi}\Gamma\left( 1+4 \Delta N^{2}\right) }{\Gamma\left( \frac{1}{2}%
+4\Delta N^{2}\right) }.  \label{First_upper_bound}
\end{equation}

It is easy to see, using the asymptotic expansions of Bessel and Gamma
functions, that for large $\Delta N$ the right hand side of this inequality
coincides with the \textit{lower} bound for $S_{N}^{\left( 2\right) }$ given
in \cite{KieferUnanyan1,KieferUnanyan2}:
\begin{equation}
S_{N}^{\left( 2\right) } \geq 2 \Delta N^{2} -\ln\left[ I_{0}\left( 2 \Delta
N^{2}\right) \right] \, \xrightarrow[\Delta N > 1] \, \ln\left( 2\sqrt{\pi}%
\Delta N\right).  \label{our_bound}
\end{equation}
Hence, $S_{N\text{ }}^{\left( 2\right) }$ can actually be
\emph{approximated} as 
\begin{equation}
S_{N\text{ }}^{\left( 2\right) }\approx\ln\left( 2\sqrt{\pi}\Delta N\right) 
\label{number_enropy_large_delta_N}
\end{equation}
for large values of $\Delta N$.

We note that the bound (\ref{First_upper_bound}) is much better than the
modified version of Shannon's inequality \cite{Cover1991} for discrete
variables 
\begin{equation}
S_{N}^{\left( 2\right) }\leq\ln\sqrt{2\pi e\left( \Delta N^{2}+\frac{1}{12}%
\right) },  \label{Shannon}
\end{equation}
which becomes $S_{N}^{\left( 2\right) }=0\leq\ln\sqrt{\frac{2\pi e}{12}}$ at 
$\Delta N\rightarrow0$ (i.e., it reduces to a trivial one). Despite being
a sharp bound on $S_{N}^{\left( 2\right) }$ for large $\Delta N,$ a
comparison with the lower bound (\ref{our_bound}) shows 
(\ref{First_upper_bound}) is not tight for small $\Delta N$. We therefore now
derive another upper bound on $S_{N}^{\left( 2\right) }$ for small $\Delta
N^2\leq\frac{1}{2}$.

Since $S_{N}^{\left( 2\right) }$ does not account for the different
configurations of particles, an obvious upper bound on $S_{N}^{\left(
2\right) }$ is given by the total Rényi entropy 
\begin{equation}
S_{N}^{\left( 2\right) }\leq S^{\left( 2\right) }=-\mathtt{tr}\ln \left( 1-%
\mathbf{G}\right) \leq -\ln \left( 1-\mathtt{tr}\mathbf{G}\right) =-\ln
\left( 1-2\Delta N^{2}\right)  \label{obvious_bound}
\end{equation}
which holds for $\Delta N^{2}\leq \frac{1}{2}.$

By combining this inequality with (\ref{First_upper_bound}) we arrive at 
\begin{equation}
S_{N}^{\left( 2\right) }\leq\left\{ 
\begin{array}{c}
-\ln\left( 1-2\Delta N^{2}\right) \text{ \ \ if }\Delta N\leq\frac{1}{2} \\ 
\ln\frac{\sqrt{\pi}\Gamma\left( 1+4\Delta N^{2}\right) }{\Gamma\left( \frac{1%
}{2}+4\Delta N^{2}\right) }\text{ \ if }\Delta N>\frac{1}{2}%
\end{array}
\right. .  \label{final_upper_bound}
\end{equation}
We note that this bound is tight for small as well as for large $\Delta N$.

\paragraph{Lower bound}

As shown in Ref.~\cite{KieferUnanyan1}, the inequality
(\ref{our_bound})---providing a lower bound for $S_{N}^{(2)}$---is
valid for any non-interacting fermion system in any dimension. We now show
that this lower bound can be further improved. To this end we show that
\begin{equation}
S_{N}^{(2)}=-\ln \sum_{n}p(n)^{2}\geq \Phi \left(2\Delta N^{2},S^{\left(
2\right) }\right) ,  \label{upper_bound} 
\end{equation}
where
\begin{equation}
\Phi \left(2\Delta N^{2},S^{\left( 2\right) }\right) =-\ln \left\{ \frac{%
e^{-2\Delta N^{2}}}{2}\bigl[I_{0}\left( 2\Delta N^{2}\right) +L_{0}\left(
2\Delta N^{2}\right) \bigr]+\frac{e^{-S^{\left( 2\right) }}}{2}\left[
I_{0}\left( S^{\left( 2\right) }\right) -L_{0}\left( S^{\left( 2\right)
}\right) \right] \right\}.\label{Phi}
\end{equation}
Here $L_{0}\left( x\right) $ is the modified Struve function. Since $%
S^{\left( 2\right) }\geq 2\Delta N^{2}$ (which follows from the obvious
inequality $\ln (1-x)\leq -x$, for $0\leq x\leq 1$ ), and $\frac{\exp \left(
-x\right) }{2}\left[ I_{0}\left( x\right) -L_{0}\left( x\right) \right] $ being
a monotonously decreasing function, we see that this bound is better than our
previous bound derived in \cite{KieferUnanyan1} which was based on the inequality 
\begin{equation}
\sum_{n}p(n)^{2}\leq \exp \left( -2\Delta N^{2}\right) I_{0}\left( 2\Delta
N^{2}\right) .  \label{our_results}
\end{equation}
In order to proof inequality (\ref{upper_bound}), we split the integral 
\begin{equation}
\sum_{n=0}^{\infty }p(n)^{2}=\frac{2}{\pi }\int_{0}^{\pi /2}\!\!\!d\theta
\,\exp \left[ \mathtt{tr}\ln \left( \mathbf{1}-2\mathbf{G}\sin ^{2}\theta
\right) \right] =U_{1}+U_{2}  \notag
\end{equation}
into two parts $U_{1}$ and $U_{2}$, where 
\begin{eqnarray}
U_{1} &=&\frac{2}{\pi }\int_{0}^{\frac{\pi }{4}}\!\!\!d\theta \,\exp \left[ 
\mathtt{tr}\ln \left( \mathbf{1}-2\mathbf{G}\sin ^{2}\theta \right) \right] ,
\\
U_{2} &=&\frac{2}{\pi }\int_{0}^{\frac{\pi }{4}}\!\!\!d\theta \,\exp \left[ 
\mathtt{tr}\ln \left( \mathbf{1}-2\mathbf{G}\cos ^{2}\theta \right) \right] .
\label{large_fluctuation}
\end{eqnarray}
The first integral can be bounded from above by using the
arithmetic-geometric inequality and the integral can then be calculated
elementary in terms of the modified Bessel and Struve functions of the first
kind resulting in 
\begin{equation}
U_{1}\leq \frac{\exp \left( -2\Delta N^{2}\right) }{2}\left[ I_{0}\left(
2\Delta N^{2}\right) +L_{0}\left( 2\Delta N^{2}\right) \right] .
\label{first_Integral}
\end{equation}
Using (\ref{log_Inequality}) and the fact that $\cos ^{2}\theta \geq \frac{1%
}{2}$ for $0\leq \theta \leq $ $\frac{\pi }{4},$ we find for the second
integral 
\begin{eqnarray}
U_{2} &\leq &\frac{2}{\pi }\int_{0}^{\frac{\pi }{4}}\!\!\!d\theta \,\exp %
\left[ 2\cos ^{2}\theta \,\mathtt{tr}\,\ln \left( \mathbf{1}-\mathbf{G}%
\right) \right] =\frac{2}{\pi }\int_{0}^{\frac{\pi }{4}}\!\!\!d\theta \,\exp %
\left[ -2S^{\left( 2\right) }\cos ^{2}\theta \right]   \notag \\
&=&\frac{1}{2}\exp \left( -S^{\left( 2\right) }\right) \left[ I_{0}\left(
S^{\left( 2\right) }\right) -L_{0}\left( S^{\left( 2\right) }\right) \right]
.  \label{second_integral}
\end{eqnarray}
Combining Eq.~(\ref{second_integral}) with Eq.(\ref{first_Integral})
one obtains expressions (\ref{upper_bound}) and (\ref{Phi}). The
derived lower bound depends on $\Delta N$ and $S^{\left( 2\right)
}$. Making use, furthermore, of either $\Delta N^{2}\geq S^{(2)}/(4\ln
(2))$ \cite{Klich2006,Muth2011,KieferUnanyan1} for large $\Delta N$
or $2 \Delta N^2 \ge 1- e^{-S(2)}$ (see Eq.~(\ref{obvious_bound})) for
small $\Delta N$, we arrive at the following connection between the
entropies
\begin{equation}
S_{N}^{\left( 2\right) }\geq \left\{ 
\begin{array}{c}
\Phi \left( 1-e^{-S^{\left( 2\right) }},S^{\left( 2\right) }\right) \text{ \
if \ }S^{\left( 2\right) }\leq \ln 2, \\ 
\Phi \left( \frac{S^{\left( 2\right) }}{2\ln 2},S^{\left( 2\right) }\right) 
\text{ \ if \ }S^{\left( 2\right) }>\ln 2%
\end{array}%
\right. .  \label{Lower_Bound_Entropies}
\end{equation}

\begin{figure}[tbp]
\includegraphics[width=1\textwidth]{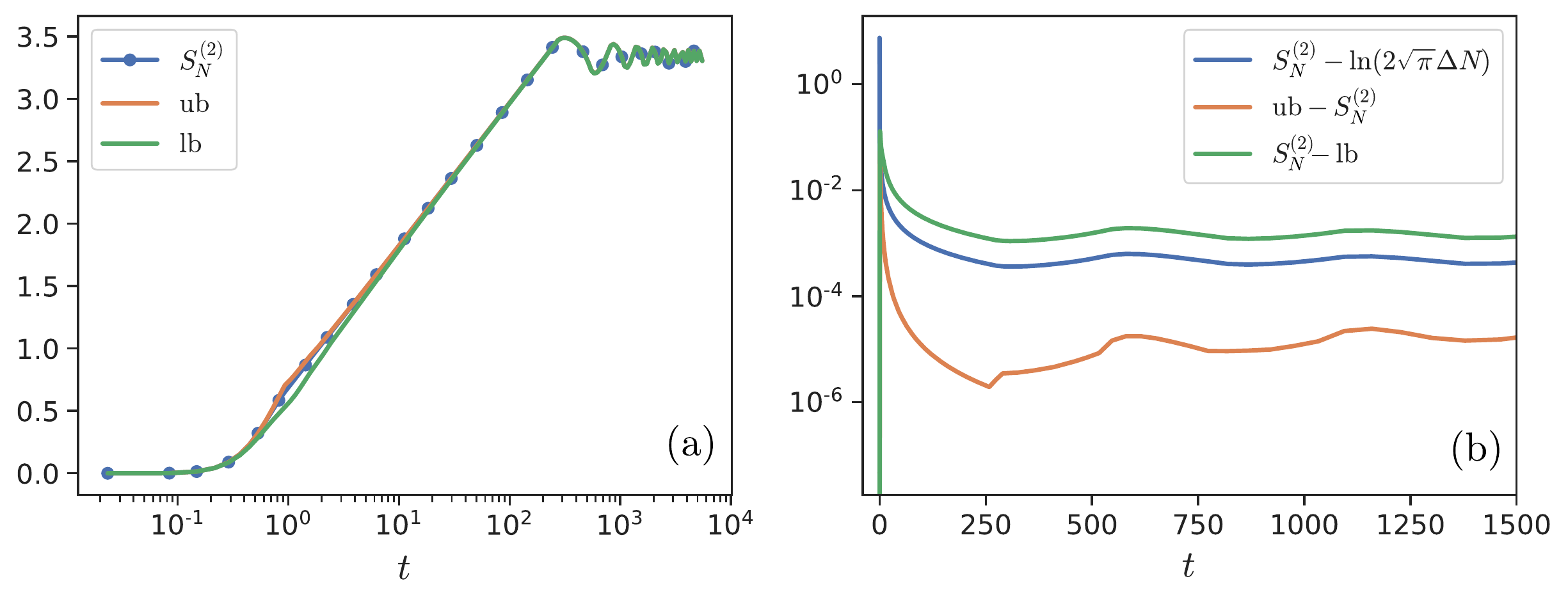}
\caption{(a) $S^{(2)}_N$ for the model \eqref{Ham} with $V=D=0$ and $L=1024$ starting from a charge-density wave inital state compared to
the upper bound (ub) (\protect\ref{final_upper_bound}) and the lower
bound (lb) (\protect\ref{upper_bound}). The bounds are quite tight as
shown by the differences between $S^{(2)}_N$ and the two bounds in
panel (b). Also $S^{(2)}_N$ can be well approximated by
$\ln(2\sqrt{\pi}\Delta N)$.}
\label{Fig6}
\end{figure}

We now present numerical checks for the quality of the derived bounds
and estimates for fermionic Gaussian models with and without
disorder. In Fig.~\ref{Fig6}(a), the time evolution of $S^{(2)}_N$ for
the Hamiltonian \eqref{Ham} with $V=D=0$ is shown.
%
%
The upper and lower bounds are very tight in this case. In
Fig.~\ref{Fig6}(b) it is shown that the second number R\'enyi entropy
is closely related to the particle number fluctuations. At long times,
the difference between the two decays exponentially before reaching a
lower limit due to the saturation of both quantities in a finite
system.

\begin{figure}[tbp]
\includegraphics[width=1\textwidth]{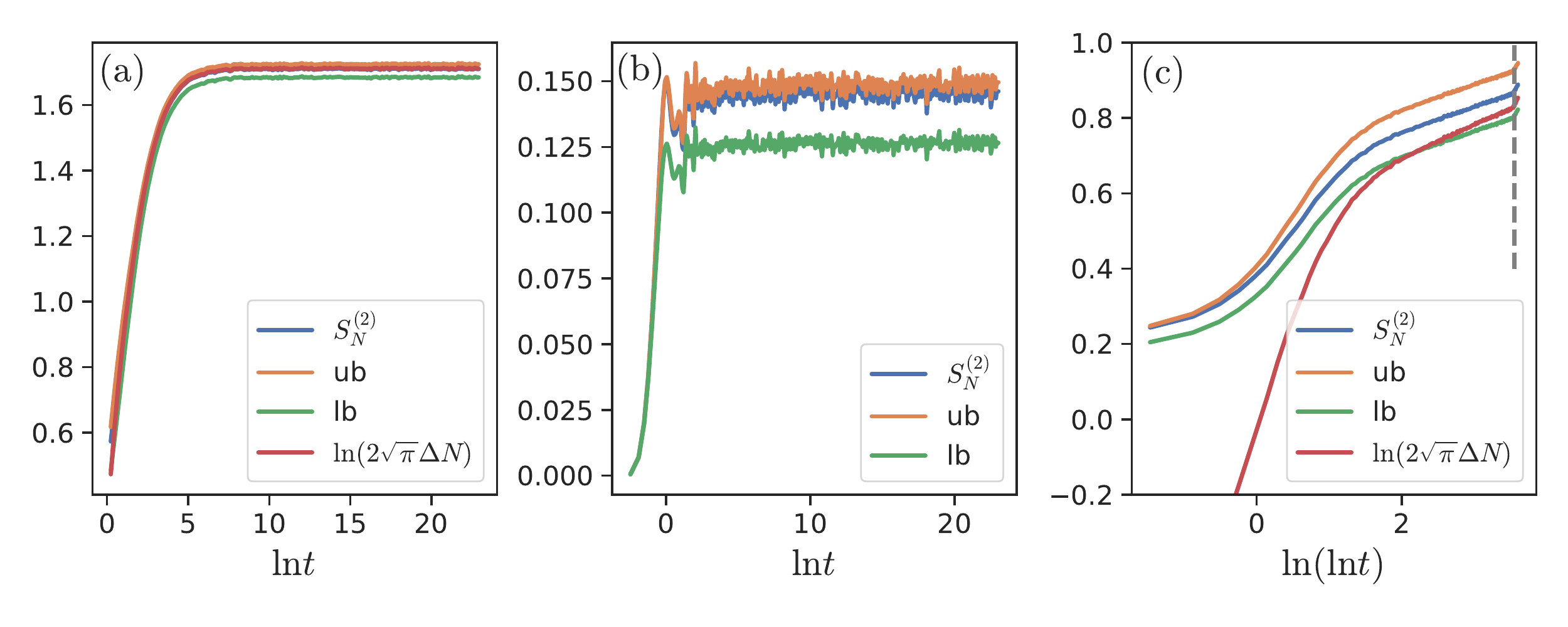} 
\caption{$S^{(2)}_N$ for the Anderson localized case, Eq.~\eqref{Ham} with $V=0$, 
for (a) $L=512$ and weak disorder, and (b) $L=128$ and strong
disorder. Averages over 2000 disorder realizations starting from a
random half-filled product state are shown. (c) Free fermions with the
random hopping amplitude $J_j$ drawn from a box distribution, the
so-called off-diagonal disorder (ODD) case, for $L=1024$ and averaged
over 20,000 disorder realizations starting from a random half-filled
product state. Shown are the upper bound (ub) corresponding to
Eq.~(\protect\ref{final_upper_bound}) and the lower bound (lb) given
by Eq.~(\protect\ref{upper_bound}). For previous results where other
bounds were used to constrain $S^{(2)}$ by $S_N$, see
Ref.~\protect\cite{KieferUnanyan1}. }
\label{Fig7}
\end{figure}

Next, we consider free fermionic systems with potential disorder (Anderson case) and off-diagonal
disorder. For the Anderson case, shown in Fig.~\ref{Fig7}(a,b), tight bounds can
be obtained both for weak and strong disorder. 
%
%
In addition, we also present in Fig.~\ref{Fig7}(c) results for the
Hamiltonian \eqref{Ham} with $V=D=0$ and with the hopping amplitude
$J$ replaced by a position dependent hopping amplitude $J_j$ which is
drawn from a box distribution. In this so-called off-diagonal disorder
(ODD) case, the entanglement entropy increases as $S^{(2)}\sim\ln \ln
t$ while the number entropy scales as $S_N^{(2)}\sim\ln\ln\ln t$
\cite{ZhaoAndraschkoSirker,KieferUnanyan1}.

\subsubsection{Free Bosons}

In this section we derive, for completeness, upper and lower bounds on the
second R\'enyi number entropy for a Gaussian bosonic state.

\paragraph{Upper bound}

The generating function $\chi\left( \theta\right) $ in the bosonic case can
be written as
\begin{equation}
\chi\left( \theta\right) =\det\left[ \mathbf{1}-\left( e^{-i\theta }-1\right) 
\frac{\mathbf{1}}{e^{\mathbf{C}}-\mathbf{1}}\right] ^{-1}.  \label{Bosonic}
\end{equation}
For bosons the matrix $\mathbf{C}$ is positive definite. Making use of
Parseval's theorem, one then finds for the number purity 
\begin{equation}
\sum_{n=0}^{\infty} p(n)^{2}=\frac{1}{2\pi}\int_{-\pi}^{\pi}\!\!\! d\theta\, 
\frac{1}{\det\left( \mathbf{1}+\mathbf{W}\left( 1-\cos\theta\right) \right) },
\qquad \mbox{where}\quad \mathbf{W}=\frac{2e^{\mathbf{C}}}{\left( e^{\mathbf{C%
}}-1\right) ^{2}}.  \label{Purity_bosons}
\end{equation}
%
The steps for obtaining bounds for $S_{N}^{\left( 2\right) }$ are the same
as in the case of free fermions. We apply the arithmetic-geometric
inequality to get an upper bound on $\det\left( \mathbf{1}+\mathbf{W}\left(
1-\cos\theta\right) \right) $, which then yields 
\begin{equation}
\sum_{n=0}^{\infty} p(n)^{2}\geq\exp\left( -\mathtt{tr}\mathbf{W}\right)
I_{0}\left( \mathtt{tr}\mathbf{W}\right) .  \label{firstUpper}
\end{equation}
Furthermore, using Eq.~(\ref{fluctuations}), one can show that $\mathtt{tr}%
\mathbf{W}$ gives the fluctuations of the total particle number, i.e.~$%
\mathtt{tr}\mathbf{W}=2\Delta N^{2}$. With inequality (\ref{firstUpper}) we
arrive at the following upper bound 
\begin{equation}
S_{N}^{\left( 2\right) }=-\ln\left( \sum_{n=0}^{\infty} p(n)^{2}\right)
\leq-\ln\left[ \exp\left( -2\Delta N^{2}\right) I_{0}\left( 2\Delta
N^{2}\right) \right] .  \label{upper_Bound}
\end{equation}
We see that this upper bound on $S_{N}^{\left( 2\right) }$ for bosons
coincides with the fermionic lower bound (\ref{our_bound}) in terms of the
particle number fluctuations.

\paragraph{Lower Bound}

By making use of the inequality $\ln\left( 1+x\right) \geq\frac{\ln\left(
1+a\right) }{a}x,$ \ \ $0\leq x\leq a$, we have 
\begin{eqnarray}
\sum_{n=0}^{\infty} p(n)^{2}&=&\frac{1}{\pi}\int_{0}^{\pi}\!\!\! d\theta\,
\exp\left( -\mathtt{tr\ln}\left[ 1+2\mathbf{W}\sin^{2}\frac{\theta}{2}\right]
\right)  \label{lower} \\
&\leq&\int_{0}^{\pi}\!\!\! d\theta\, \exp\left( -\sin^{2}\frac{\theta}{2}%
\mathtt{tr\ln}\left[ 1+2\mathbf{W}\right] \right) =\exp\left( -S^{\left(
2\right) }\right) I_{0}\left( S^{\left( 2\right) }\right),  \notag
\end{eqnarray}
where in the last line we have used the identity %
$S^{\left( 2\right) }=\frac{1}{2}\mathtt{tr}\ln\left( \mathbf{1}+2\mathbf{W}%
\right)$. 
%
Hence, we arrive at the lower bound 
\begin{equation}
S_{N}^{\left( 2\right) }\geq-\ln\left[ \exp\left( -S^{\left( 2\right)
}\right) I_{0}\left( S^{\left( 2\right) }\right) \right] \, .
\label{lower_bound_S2}
\end{equation}


\subsection{Relation between number entropy and number fluctuations for
interacting systems}

\begin{figure}[tbp]
\includegraphics[width=1\textwidth]{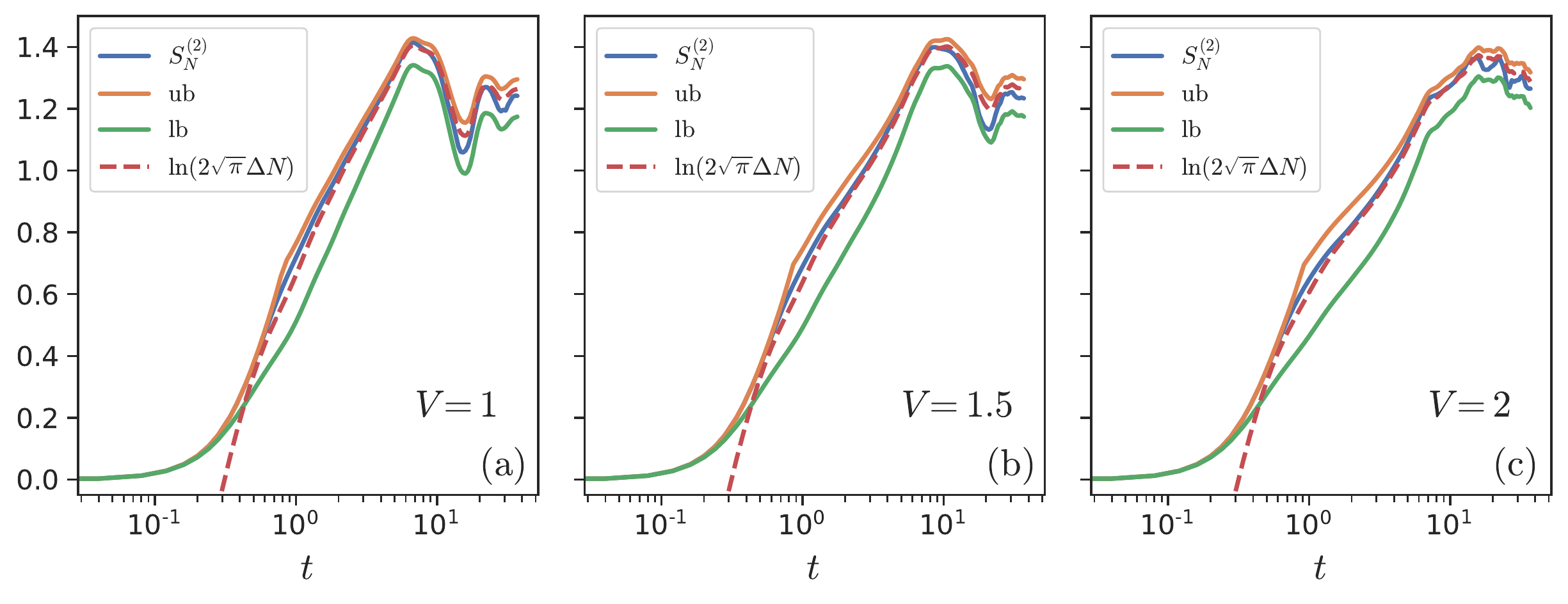}
\caption{Results for the dynamics of the number entropy $S_N^{(2)}$ for $L=24$ when
quenching the $t-V$ model from a charge-density wave initial state compared to
the upper bound (ub) (\protect\ref{final_upper_bound}) and the lower bound
(lb) (\protect\ref{our_bound}) for different interaction strengths but
without disorder. Furthermore, we can see that $S_N^{(2)}\approx \mathrm{ln}%
(2\protect\sqrt{\protect\pi}\Delta N)$ holds even in the interacting case. }
\label{Fig8}
\end{figure}

In the previous section, we have established a tight relation between the
R\'enyi number entropy $S_N^{(2)}$ and the number fluctuations $\Delta N$ in
non-interacting, i.e. Gaussian systems, expressed by the lower and upper
bounds, Eqs.~\eqref{our_bound} and \eqref{final_upper_bound}, respectively,
\begin{equation}
2 \Delta N^{2} -\ln\left[ I_{0}\left( 2 \Delta N^{2}\right) \right] \leq
S_{N}^{\left( 2\right) }\leq\left\{ 
\begin{array}{c}
-\ln\left( 1-2\Delta N^{2}\right) \text{ \ \ if }\Delta N\leq\frac{1}{2} \\ 
\ln\frac{\sqrt{\pi}\Gamma\left( 1+4\Delta N^{2}\right) }{\Gamma\left( \frac{1%
}{2}+4\Delta N^{2}\right) }\text{ \ if }\Delta N>\frac{1}{2}%
\end{array}
\right. .  \label{lower-upper}
\end{equation}
A natural question that arises is whether these bounds also hold for
interacting systems. In Fig. \ref{Fig8}, we show the number entropy as
well as the lower bound (lb), and the upper bound (ub) for the $t-V$
model without disorder. We recognize that the two bounds as well as
the estimate, $S_N^{(2)}\approx \ln\bigl(2\sqrt{\pi} \Delta N\bigr)$
for large $\Delta N$, hold true even in the interacting case. Since
both quantities $S^{(2)}_N$ and $\Delta N$ depend on the same
probability distribution $p(n)$, this might not be too surprising. It
does show, however, that an unlimited growth of the R\'enyi number
entropy implies a corresponding growth of number
fluctuations. Numerical results for $\Delta N$ in the putative MBL
phase are discussed in \ref{AppB}.

%
%
%
%


\subsection{Renormalized bounds for the number entropy in terms of the entanglement
entropy for interacting systems}


The bounds for the number entropy derived above for Gaussian,
i.e.~non-interacting, systems also establish a relation to the
entanglement entropy
\begin{equation}
S^{(2)}_N\sim \ln S^{(2)}.  \label{entropies}
\end{equation}
This relation is consistent with our observation that $S^{(2)}\sim\ln
t$ and $S^{(2)}_N\sim\ln\ln t$ in the putative many-body localized
phase. In the following, we show that the lower bound for $S_N^{(2)}$
in terms of $S^{(2)}$, which leads to relation \eqref{entropies},
indeed appears to hold also for the interacting t-V model including in
the MBL phase with some renormalization. This provides further
evidence that the particle fluctuations are not bounded.

\begin{figure}[h]
\includegraphics[width=1\textwidth]{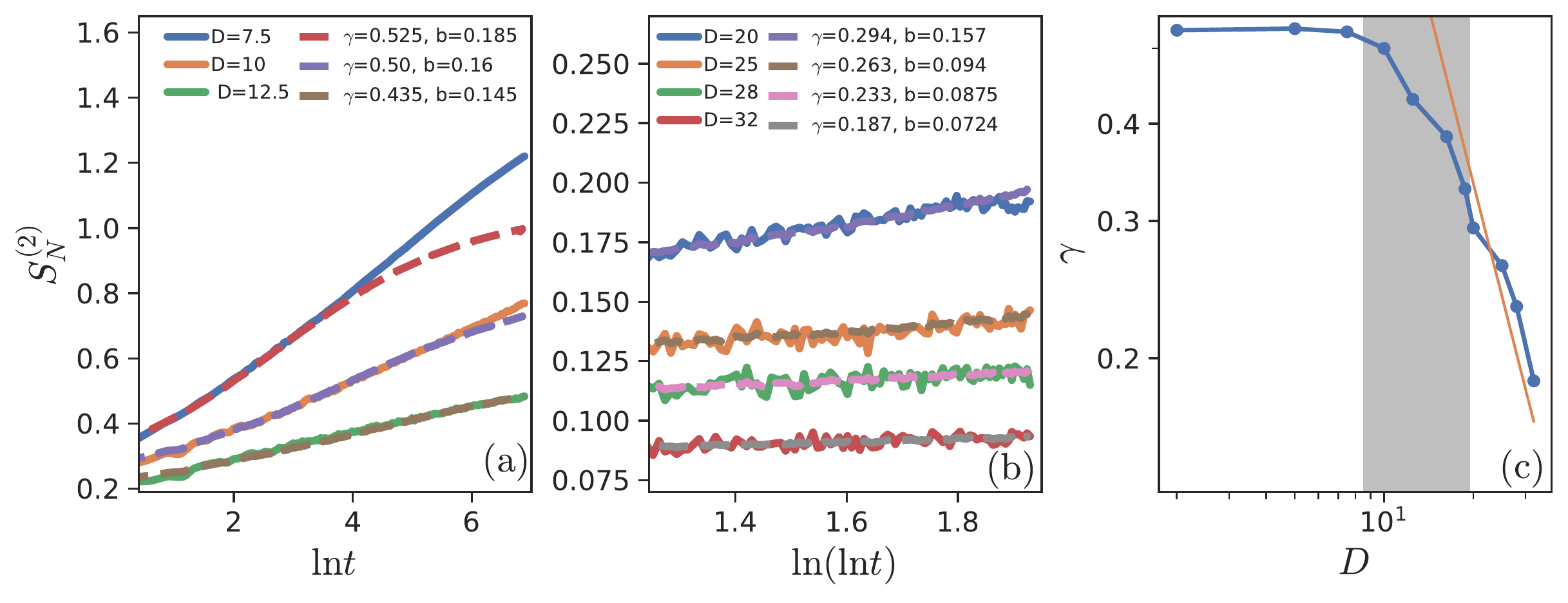}
\caption{Dynamics of $S_N^{(2)}$ for $L=24$ compared to the renormalized lower bound (\protect\ref%
{upper_bound_dS2}) (dashed lines) for (a) $D<D_c$, and (b)
$D>D_c$. The renormalized lower bounds are shifted by a constant $b$
to demonstrate that the bound shows the same scaling as
$S^{(2)}_N$. (a-b) are partially based on data already used in
Ref.~\protect\cite{KieferUnanyan2}. (c) Renormalization factor
$\protect\gamma$ as a function of disorder. For $D<D_c$,
$\protect\gamma$ remains close to $\protect\gamma\sim 0.5$ but decays
approximately like a power-law for $D> D_c$. Here we have averaged over 1500
disorder realizations for $D\leq 28$ and 2000 for $D\geq 28$, starting
from a random half-filled product state.}
\label{Fig9}
\end{figure}

To this end, we introduce a renormalization prefactor $\gamma \leq 1$
in (\ref{Lower_Bound_Entropies}), see also \cite{KieferUnanyan2}
\begin{equation}
S_{N}^{(2)}\geq \left\{ 
\begin{array}{c}
\gamma \Phi \left( 1-e^{-S^{\left( 2\right) }},S^{\left( 2\right) }\right) 
\text{ \ if \ }S^{\left( 2\right) }\leq \ln 2, \\ 
\gamma \Phi \left( \frac{S^{\left( 2\right) }}{2\ln 2},S^{\left( 2\right)
}\right) \text{ \ if \ }S^{\left( 2\right) }>\ln 2%
\end{array}%
\right. .  \label{upper_bound_dS2}
\end{equation}
The factor $\gamma $ is needed as the lower bound will be broken at some
point in time otherwise. We compare $S_{N}^{(2)}$ to the renormalized lower
bound, see Fig.~\ref{Fig9}, for the disordered $t-V$ model. We find that this
bound describes the data very well both for $D<D_{c}$ and for $D>D_{c}$. As
can be seen in Fig.~\ref{Fig9}(c), the prefactor $\gamma $ does depend smoothly on the
disorder strength. We find, in particular, that $\gamma $ is almost constant for $D<D_{c}$ 
and falls off approximately like a power law above $D_c$.


\section{Conclusions}

\label{Sec_Concl} 
In conclusion, we have provided strong arguments why the putative MBL
phase in the disordered one-dimensional t-V model (isotropic
Heisenberg chain) does not appear to be truly localized. Our arguments
are based on the numerical evaluation of the time evolution after a
quantum quench, the results of which are summarized in Table
\ref{Tab1}. In the interacting case, our simulations have been carried
out in systems up to lengths of $L=24$. We therefore obviously cannot
exclude scenarios where the behavior of the particle fluctuations
qualitatively changes for larger system sizes and longer times than
numerically accessible. We note, however, that if one is going to
dismiss the results presented here as valid arguments against
localization, one should then also dismiss any arguments in favor of
localization coming from numerical studies of small systems. This
would ultimately mean that we currently cannot numerically study
whether or not MBL phases exist.
\begin{table}[ht!] 
\begin{center}
\begin{tabular}{c||c|c|c|c}
\textbf{Phase} & \textbf{disorder} & \textbf{interaction} & $S^{(2)}$ & $%
S_N^{(2)}$ \\ \hline\hline
AL & potential, $D\neq 0$ & non-interacting & $\sim\mbox{const}$ & $\sim%
\mbox{const}$ \\ \hline
ODD & off-diagonal & non-interacting & $\sim\ln\ln t$ & $\sim\ln\ln\ln t$ \\ 
\hline
MBL & potential, $D>D_c$ & interacting & $\sim\ln t$ & $\sim\ln\ln t$ \\ 
\hline
clean & no-disorder & non-int. \& int. & $\sim t$ & $\sim\ln t$%
\end{tabular}%
\end{center}
\caption{Asymptotic growth of R\'enyi entropy $S^{(2)}$ and R\'enyi number
entropy $S_N^{(2)}$ following a quantum quench for various phases of the t-V
model. The relation $S_N^{(2)}\sim\ln S^{(2)}$ appears to hold in all of them.}
\label{Tab1}
\end{table}

If we assume that we can learn something about putative MBL phases in
one dimension from studies of small systems, then it appears to be
clear that the evidence now points towards an absence of true
localization. Our original arguments made in
Ref.~\cite{KieferUnanyan2} were based on the observation that the
number entropy grows as $S_N\sim\ln\ln t$ and that this growth is
consistent with the relation $S_N\sim \ln S$ thus pointing to an
unbounded growth of the number entropy.

In the present article, we have tried to address possible criticisms
of this interpretation of the data. First of all, we have shown that
all the data for disorder strengths close to the assumed ergodic-MBL
phase transition at $D=D_c$ up to disorder strengths of more than
$2D_c$ are consistently described by $S_N=\frac{\nu}{2}\ln\ln t$ and
that the prefactor $\nu$ shows a power-law dependence on disorder
strength. Furthermore, calculating the median of the number entropy we
showed that the observed scaling represents typical behavior and is
not the result of rare disorder realizations. This indicates that the
increase of the number entropy is a generic feature of the MBL phase
and not restricted to disorder strengths close to the phase
transition. Second, we have demonstrated that the deviation time
$t_d$, where the finite-size data start to deviate from the double
logarithmic fit, scales as $t_d\sim \exp(L/\ell)$ with a
characteristic length scale $\ell \sim (D-D_c)^{-0.5}$, which is well
defined only for $D>D_c$. We have shown that all the data for $t_d$,
obtained for various different system sizes and disorder strengths,
show an excellent scaling collapse. The double logarithmic scaling of
the number entropy in time therefore does not appear to be transient
but rather indicative of the thermodynamic limit. Lastly, we have
shown that the observed increase of the number entropy cannot be
explained by the fluctuations of a small number of particles initially
situated near the cut between the two subsystems. Instead, we have
found that the particle number distribution $p(n)$ as a whole becomes
wider over time. In particular, large particle number fluctuations are
becoming increasingly more likely. This is most clearly seen in the
truncated Hartley number entropy $S_H$ which counts the number of
particle configurations $n$ with $p(n)>p_c$ where $p_c$ is some
cutoff. We have shown that $S_H\sim\ln\ln t$ for the putative MBL
phase while $S_H$ saturates quickly in the Anderson case.

To shed some more light on the relation between the growth of the
entanglement and number entropies, we have considered free fermionic
and bosonic systems. In both cases, we have been able to derive strict
upper and lower bounds and have numerically shown that these bounds
can be very tight in specific cases. We have argued that these bounds,
with renormalized coefficients, also hold in the putative MBL phase
further supporting the conclusion that $S_N\sim \ln S$, i.e., the
unbounded logarithmic growth of the entanglement entropy is
accompanied by an unbounded double logarithmic growth of the number
entropy. In light of these findings, we believe that the very notion
of many-body localization in one dimension needs to be reconsidered.

Finally, we would like to emphasize again that we have not made any
statements about the putative ergodic-MBL phase transition so far. We
can think of at least two scenarios for the phase diagram of the t-V
model which are consistent with our data: (1) One possibility would be
that there is no phase transition but rather a crossover with the
particle dynamics becoming slower and slower with increasing
disorder. Such a scenario would appear to be consistent with recent
results in
Refs.~\cite{SuntajsBonca,SuntajsBonca2,SelsPolkovnikov,SelsPolkovnikov2}. We
want to point out, in particular, that $S \sim
\ln(t)\sim(t^\nu-1)/\nu$ and $S_N\sim\ln\ln(t) \sim
\ln[(t^\nu-1)/\nu]$ for $\nu$ small, i.e., a crossover where $\nu\to
0$ for $D\to\infty$ would be very difficult to distinguish numerically
from a true change in scaling at a phase transition. We note that in
this case sub-diffusive transport would prevail at very long times
$t^\nu\gg 1$ for a small but finite $\nu$. (2) A second possibility
might be that there is indeed a phase transition at a critical
disorder strength $D_c$ with the system for $D>D_c$ having both
extended and localized states. In this regard, we note that the
observed scaling of the entanglement entropy and the number entropy
appears to be the same as the one recently found right at the phase
transition in the three-dimensional Anderson model
\cite{ZhaoSirker2020}. I.e., the phase for $D>D_c$ could be more akin
to an extended critical phase. If such a transition would be at all
possible and what the nature of such a transition would be is,
however, unclear.

It would also be of interest to conduct similar studies of the R\'enyi
number entropies and particle fluctuations in interacting disordered
many-body systems in higher dimensions. We note that in this case the
instability of MBL in the thermodynamic limit appears to be far less
controversial although the situation is far from completely settled
either \cite{ChoiHild,WahlPal,Grozdanov2015,Grozdanov2016}.


\section*{Acknowledgement}

J.~S. acknowledges support by the Natural Sciences and Engineering
Research Council (NSERC, Canada) and by the Deutsche Forschungsgemeinschaft
(DFG) via Research Unit FOR 2316. M.~K., R.~U. and M.~F. acknowledge financial
support from the Deutsche Forschungsgemeinschaft (DFG) via SFB TR185,
project number 277625399. The simulations were (partly) executed on the high
performance cluster "Elwetritsch" at the University of Kaiserslautern
which is part of the "Alliance of High Performance Computing
Rheinland-Pfalz" (AHRP). We kindly acknowledge the support of the
RHRK.

\appendix

\section{Number entropy for different realizations}
\label{AppA}
\setcounter{figure}{0}
An important question for the interpretation of our results is whether
or not the observed scaling of the number entropy $S_N\sim\ln\ln t$ is
related to rare configurations and rare initial states. In
Sec.~\ref{Close}, we have tried to answer this question by comparing
the average with the median entropy and found the same scaling for
both quantities. Here we want to go one step further and consider the
number entropy for samples sorted into ten bins according to the
magnitude of $S_N(t)$ at each time step and averaged over each bin
individually. The result for two disorder strengths is shown in
Fig.~\ref{Fig1AppA}.
\begin{figure}[h!]
\includegraphics[width=\textwidth]{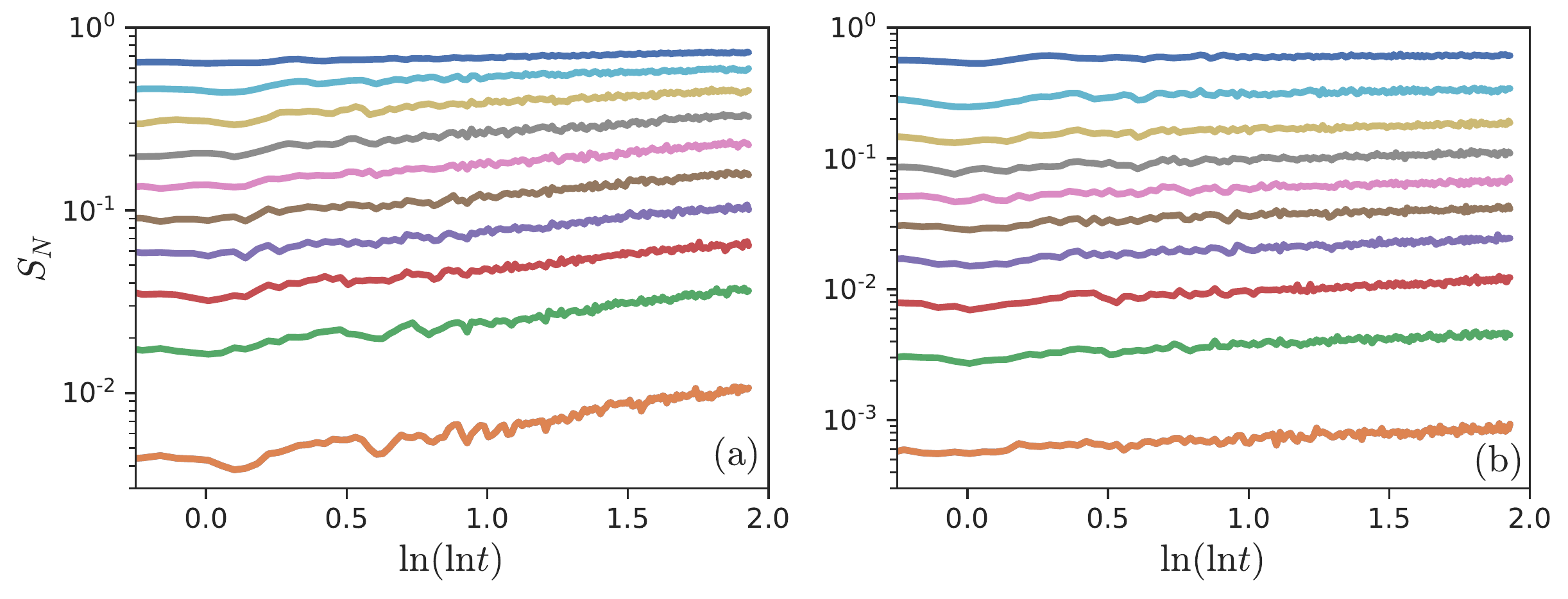}
\caption{Number entropies for (a) $D=20$ and (b) $D=32$. 
Here we have sorted all samples into ten bins at each time step and
averaged over each bin individually.}
\label{Fig1AppA}
\end{figure}
We find that almost all the bins show a scaling $S_N\sim\ln\ln t$. For
the bin containing the samples with the largest values of $S_N$, the
number entropy rises quickly to values close to their finite-size
saturation values, i.e., for the rare samples which do contain regions
with little disorder the saturation value is reached even quicker in
time. This further supports the notion that the double logarithmic
scaling in time is the typical behavior for all $D>D_c$ and is not
related to any special rare configurations.

\section{Number fluctuations}
\label{AppB}
\setcounter{figure}{0}
We have found it useful to concentrate mostly on the R\'enyi number
entropies instead of studying the particle number fluctuations $\Delta
N$ in a partition directly. The main reason to do so is that the
scaling of the number entropy can be directly related to the $S\sim\ln
t$ scaling of the entanglement entropy which is supposed to be one of
the hallmarks of the putative MBL phase. In particular, we have shown
that it is possible to derive bounds for $S_N^{(2)}$ in terms of
$S^{(2)}$ for Gaussian systems which also appear to hold for the
interacting case. Nevertheless, the R\'enyi number entropies
$S^{(\alpha)}_N$ and the particle fluctuations $\Delta N$ in a
subsystem depend of course both on the same particle distribution
function $p(n)$. We therefore expect that $\Delta N(t)$ is also
continuously growing in time. That this is indeed the case is shown in
Fig.~\ref{Fig1AppB}.
\begin{figure}[h!]
\begin{center}
\includegraphics[width=0.6\textwidth]{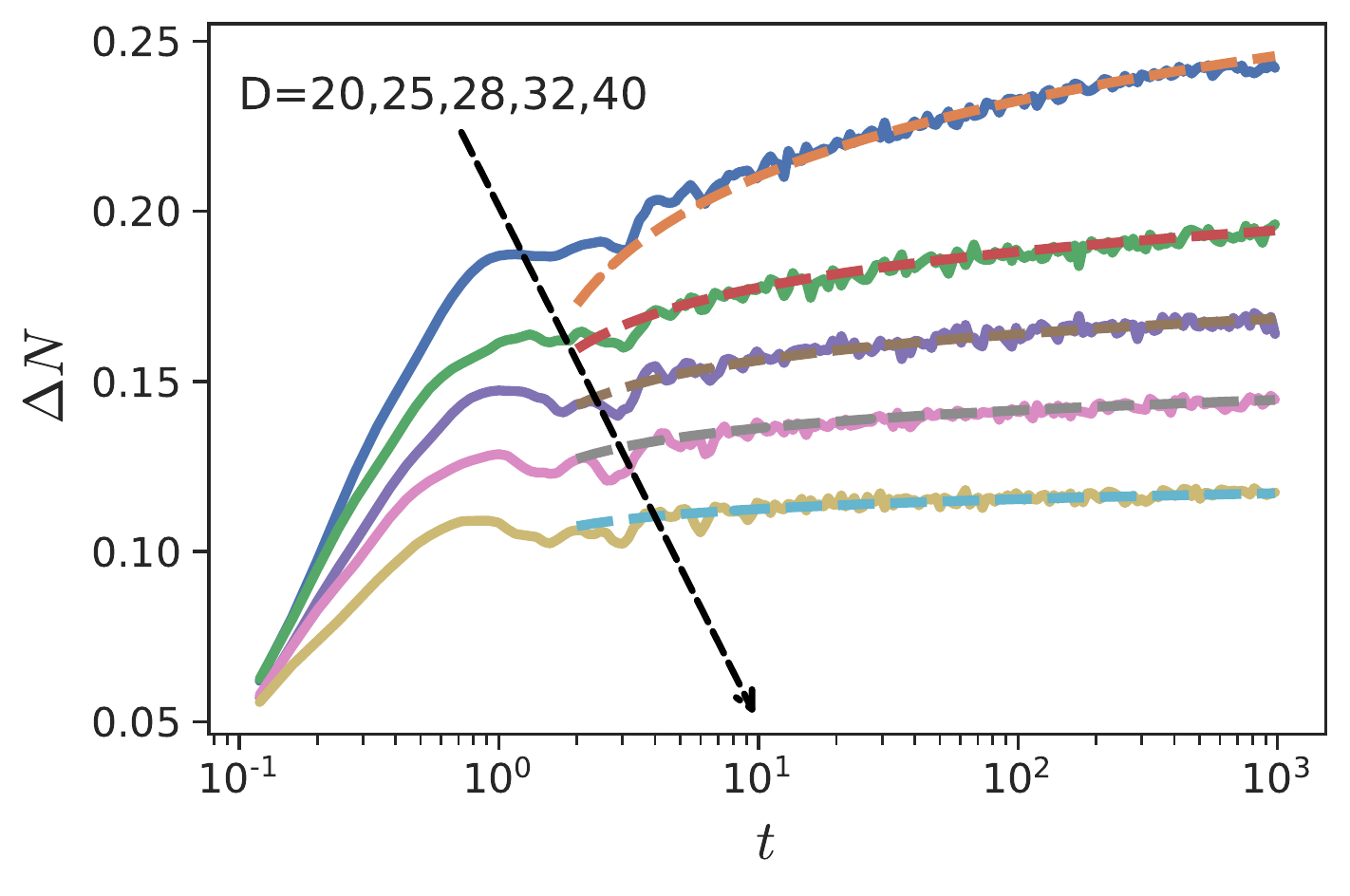}
\end{center}
\caption{Averaged particle number fluctuations $\Delta N$ in one partition as a function of time. The lines are fits $\Delta N\sim (\ln t)^\nu + b$.}
\label{Fig1AppB}
\end{figure}
We find that the data for all disorder strengths are well fitted by
$\Delta N\sim (\ln t)^\nu + b$ which is consistent with the results
for the R\'enyi number entropies presented in the main text.


\begin{thebibliography}{10}
\expandafter\ifx\csname url\endcsname\relax
  \def\url#1{\texttt{#1}}\fi
\expandafter\ifx\csname urlprefix\endcsname\relax\def\urlprefix{URL }\fi
\expandafter\ifx\csname href\endcsname\relax
  \def\href#1#2{#2} \def\path#1{#1}\fi

\bibitem{Anderson58}
P.~W. Anderson, Absence of diffusion in certain random lattices, Phys. Rev. 109
  (1958) 1492--1505.

\bibitem{AndersonLocalization}
E.~Abrahams (Ed.), 50 Years of Anderson Localization, World Scientific,
  Singapore, 2010.

\bibitem{AbrahamsAnderson}
E.~Abrahams, P.~W. Anderson, D.~C. Licciardello, T.~V. Ramakrishnan, Scaling
  theory of localization: Absence of quantum diffusion in two dimensions, Phys.
  Rev. Lett. 42 (1979) 673--676.
\newblock \href {http://dx.doi.org/10.1103/PhysRevLett.42.673}
  {\path{doi:10.1103/PhysRevLett.42.673}}.

\bibitem{EdwardsThouless}
J.~T. Edwards, D.~J. Thouless, Numerical studies of localization in disordered
  systems, J. Phys. C 5~(8) (1972) 807.

\bibitem{OganesyanHuse}
V.~Oganesyan, D.~A. Huse, Localization of interacting fermions at high
  temperature, Phys. Rev. B 75 (2007) 155111.

\bibitem{PalHuse}
A.~Pal, D.~A. Huse, Many-body localization phase transition, Phys. Rev. B 82
  (2010) 174411.

\bibitem{EnssAndraschkoSirker}
T.~Enss, F.~Andraschko, J.~Sirker, Many-body localization in infinite chains,
  Phys. Rev. B 95 (2017) 045121.

\bibitem{BaskoAleiner}
D.~M. Basko, I.~L. Aleiner, B.~L. Altshuler, Metal-insulator transition in a
  weakly interacting many-electron system with localized single-particle
  states, Ann. Phys.

\bibitem{Imbrie2016}
J.~Z. Imbrie, Diagonalization and many-body localization for a disordered
  quantum spin chain, Phys. Rev. Lett. 117 (2016) 027201.

\bibitem{Luitz1}
D.~J. Luitz, N.~Laflorencie, F.~Alet, Many-body localization edge in the
  random-field heisenberg chain, Phys. Rev. B 91 (2015) 081103.

\bibitem{ZnidaricProsen}
M.~\ifmmode \check{Z}\else \v{Z}\fi{}nidari\ifmmode~\check{c}\else \v{c}\fi{},
  T.~Prosen, P.~Prelov\ifmmode~\check{s}\else \v{s}\fi{}ek, Many-body
  localization in the heisenberg $xxz$ magnet in a random field, Phys. Rev. B
  77 (2008) 064426.

\bibitem{BardarsonPollmann}
J.~H. Bardarson, F.~Pollmann, J.~E. Moore, Unbounded growth of entanglement in
  models of many-body localization, Phys. Rev. Lett. 109 (2012) 017202.
\newblock \href {http://dx.doi.org/10.1103/PhysRevLett.109.017202}
  {\path{doi:10.1103/PhysRevLett.109.017202}}.

\bibitem{SerbynPapic}
M.~Serbyn, Z.~Papi\ifmmode~\acute{c}\else \'{c}\fi{}, D.~A. Abanin, Local
  conservation laws and the structure of the many-body localized states, Phys.
  Rev. Lett. 111 (2013) 127201.

\bibitem{HuseNandkishore}
D.~A. Huse, R.~Nandkishore, V.~Oganesyan, Phenomenology of fully
  many-body-localized systems, Phys. Rev. B 90 (2014) 174202.

\bibitem{VoskHusePRX}
R.~Vosk, D.~A. Huse, E.~Altman, Theory of the many-body localization transition
  in one-dimensional systems, Phys. Rev. X 5 (2015) 031032.

\bibitem{Goremykina2019}
A.~Goremykina, R.~Vasseur, M.~Serbyn,
  \href{https://link.aps.org/doi/10.1103/PhysRevLett.122.040601}{Analytically
  solvable renormalization group for the many-body localization transition},
  Phys. Rev. Lett. 122 (2019) 040601.
\newblock \href {http://dx.doi.org/10.1103/PhysRevLett.122.040601}
  {\path{doi:10.1103/PhysRevLett.122.040601}}.
\newline\urlprefix\url{https://link.aps.org/doi/10.1103/PhysRevLett.122.040601}

\bibitem{Dumitrescu2019}
P.~T. Dumitrescu, A.~Goremykina, S.~A. Parameswaran, M.~Serbyn, R.~Vasseur,
  \href{https://link.aps.org/doi/10.1103/PhysRevB.99.094205}{Kosterlitz-thouless
  scaling at many-body localization phase transitions}, Phys. Rev. B 99 (2019)
  094205.
\newblock \href {http://dx.doi.org/10.1103/PhysRevB.99.094205}
  {\path{doi:10.1103/PhysRevB.99.094205}}.
\newline\urlprefix\url{https://link.aps.org/doi/10.1103/PhysRevB.99.094205}

\bibitem{PotterVasseurPRX}
A.~C. Potter, R.~Vasseur, S.~A. Parameswaran, Universal properties of many-body
  delocalization transitions, Phys. Rev. X 5 (2015) 031033.

\bibitem{MorningstarHuse}
A.~Morningstar, D.~A. Huse, J.~Z. Imbrie, Many-body localization near the
  critical point, Phys. Rev. B 102 (2020) 125134.

\bibitem{SuntajsBonca}
J.~Suntajs, J.~Bonca, T.~Prosen, L.~Vidmar, Quantum chaos challenges many-body
  localization, arXiv: 1905.06345.

\bibitem{SuntajsBonca2}
J.~Suntajs, J.~Bonca, T.~Prosen, L.~Vidmar, Ergodicity breaking transition in
  finite disordered spin chains, arXiv: 2004.01719.

\bibitem{SelsPolkovnikov}
D.~Sels, A.~Polkovnikov, Dynamical obstruction to localization in a disordered
  spin chain, arXiv: 2009.04501.

\bibitem{SelsPolkovnikov2}
T.~LeBlond, D.~Sels, A.~Polkovnikov, M.~Rigol, Universality in the onset of
  quantum chaos in many-body systems, arXiv: 2012.07849.

\bibitem{KieferUnanyan1}
M.~Kiefer-Emmanouilidis, R.~Unanyan, J.~Sirker, M.~Fleischhauer, Bounds on the
  entanglement entropy by the number entropy in non-interacting fermionic
  systems, SciPost Phys. 8 (2020) 083.

\bibitem{KieferUnanyan2}
M.~Kiefer-Emmanouilidis, R.~Unanyan, M.~Fleischhauer, J.~Sirker, Evidence for
  unbounded growth of the number entropy in many-body localized phases, Phys.
  Rev. Lett. 124 (2020) 243601.

\bibitem{KieferUnanyan3}
M.~Kiefer-Emmanouilidis, R.~Unanyan, M.~Fleischhauer, J.~Sirker, Absence of
  true localization in many-body localized phases, arXiv:2010.00565.

\bibitem{AndraschkoEnssSirker}
F.~Andraschko, T.~Enss, J.~Sirker, Purification and many-body localization in
  cold atomic gases, Phys. Rev. Lett. 113 (2014) 217201.

\bibitem{Doggen2018}
E.~V.~H. Doggen, F.~Schindler, K.~S. Tikhonov, A.~D. Mirlin, T.~Neupert, D.~G.
  Polyakov, I.~V. Gornyi,
  \href{https://link.aps.org/doi/10.1103/PhysRevB.98.174202}{Many-body
  localization and delocalization in large quantum chains}, Phys. Rev. B 98
  (2018) 174202.
\newblock \href {http://dx.doi.org/10.1103/PhysRevB.98.174202}
  {\path{doi:10.1103/PhysRevB.98.174202}}.
\newline\urlprefix\url{https://link.aps.org/doi/10.1103/PhysRevB.98.174202}

\bibitem{Doggen2019}
E.~V.~H. Doggen, A.~D. Mirlin,
  \href{https://link.aps.org/doi/10.1103/PhysRevB.100.104203}{Many-body
  delocalization dynamics in long aubry-andr\'e quasiperiodic chains}, Phys.
  Rev. B 100 (2019) 104203.
\newblock \href {http://dx.doi.org/10.1103/PhysRevB.100.104203}
  {\path{doi:10.1103/PhysRevB.100.104203}}.
\newline\urlprefix\url{https://link.aps.org/doi/10.1103/PhysRevB.100.104203}

\bibitem{Abaninrecent}
D.~A. Abanin, J.~H. Bardarson, G.~de~Tomasi, S.~Gopalakrishnan, V.~Khemani,
  S.~A. Parameswaran, F.~Pollmann, A.~C. Potter, M.~Serbyn, R.~Vasseur,
  Distinguishing localization from chaos: challenges in finite-size systems,
  arXiv: 1911.04501.

\bibitem{Sierant2020}
P.~Sierant, D.~Delande, J.~Zakrzewski,
  \href{https://link.aps.org/doi/10.1103/PhysRevLett.124.186601}{Thouless time
  analysis of anderson and many-body localization transitions}, Phys. Rev.
  Lett. 124 (2020) 186601.
\newblock \href {http://dx.doi.org/10.1103/PhysRevLett.124.186601}
  {\path{doi:10.1103/PhysRevLett.124.186601}}.
\newline\urlprefix\url{https://link.aps.org/doi/10.1103/PhysRevLett.124.186601}

\bibitem{Buijsman2020}
W.~Buijsman, V.~Cheianov, V.~Gritsev,
  \href{https://link.aps.org/doi/10.1103/PhysRevE.102.042216}{Sensitivity of
  the spectral form factor to short-range level statistics}, Phys. Rev. E 102
  (2020) 042216.
\newblock \href {http://dx.doi.org/10.1103/PhysRevE.102.042216}
  {\path{doi:10.1103/PhysRevE.102.042216}}.
\newline\urlprefix\url{https://link.aps.org/doi/10.1103/PhysRevE.102.042216}

\bibitem{LuitzBarLev}
D.~J. Luitz, Y.~B. Lev,
  \href{https://link.aps.org/doi/10.1103/PhysRevB.102.100202}{Absence of slow
  particle transport in the many-body localized phase}, Phys. Rev. B 102 (2020)
  100202.
\newblock \href {http://dx.doi.org/10.1103/PhysRevB.102.100202}
  {\path{doi:10.1103/PhysRevB.102.100202}}.
\newline\urlprefix\url{https://link.aps.org/doi/10.1103/PhysRevB.102.100202}

\bibitem{KlichLevitov}
I.~Klich, L.~S. Levitov, Scaling of entanglement entropy and superselection
  rules, arXiv:0812.0006.

\bibitem{WisemanVaccaro}
H.~M. Wiseman, J.~A. Vaccaro,
  \href{https://link.aps.org/doi/10.1103/PhysRevLett.91.097902}{Entanglement of
  indistinguishable particles shared between two parties}, Phys. Rev. Lett. 91
  (2003) 097902.
\newblock \href {http://dx.doi.org/10.1103/PhysRevLett.91.097902}
  {\path{doi:10.1103/PhysRevLett.91.097902}}.
\newline\urlprefix\url{https://link.aps.org/doi/10.1103/PhysRevLett.91.097902}

\bibitem{DowlingDohertyWiseman}
M.~R. Dowling, A.~C. Doherty, H.~M. Wiseman,
  \href{https://link.aps.org/doi/10.1103/PhysRevA.73.052323}{Entanglement of
  indistinguishable particles in condensed-matter physics}, Phys. Rev. A 73
  (2006) 052323.
\newblock \href {http://dx.doi.org/10.1103/PhysRevA.73.052323}
  {\path{doi:10.1103/PhysRevA.73.052323}}.
\newline\urlprefix\url{https://link.aps.org/doi/10.1103/PhysRevA.73.052323}

\bibitem{SchuchVerstraeteCirac}
N.~Schuch, F.~Verstraete, J.~I. Cirac,
  \href{https://link.aps.org/doi/10.1103/PhysRevLett.92.087904}{Nonlocal
  resources in the presence of superselection rules}, Phys. Rev. Lett. 92
  (2004) 087904.
\newblock \href {http://dx.doi.org/10.1103/PhysRevLett.92.087904}
  {\path{doi:10.1103/PhysRevLett.92.087904}}.
\newline\urlprefix\url{https://link.aps.org/doi/10.1103/PhysRevLett.92.087904}

\bibitem{SchuchVerstraeteCirac2}
N.~Schuch, F.~Verstraete, J.~I. Cirac,
  \href{https://link.aps.org/doi/10.1103/PhysRevA.70.042310}{Quantum
  entanglement theory in the presence of superselection rules}, Phys. Rev. A 70
  (2004) 042310.
\newblock \href {http://dx.doi.org/10.1103/PhysRevA.70.042310}
  {\path{doi:10.1103/PhysRevA.70.042310}}.
\newline\urlprefix\url{https://link.aps.org/doi/10.1103/PhysRevA.70.042310}

\bibitem{SongFlindt}
H.~F. Song, C.~Flindt, S.~Rachel, I.~Klich, K.~Le~Hur,
  \href{https://link.aps.org/doi/10.1103/PhysRevB.83.161408}{Entanglement
  entropy from charge statistics: Exact relations for noninteracting many-body
  systems}, Phys. Rev. B 83 (2011) 161408.
\newblock \href {http://dx.doi.org/10.1103/PhysRevB.83.161408}
  {\path{doi:10.1103/PhysRevB.83.161408}}.
\newline\urlprefix\url{https://link.aps.org/doi/10.1103/PhysRevB.83.161408}

\bibitem{Rakovszky2019}
T.~Rakovszky, C.~W. von Keyserlingk, F.~Pollmann,
  \href{https://link.aps.org/doi/10.1103/PhysRevB.100.125139}{Entanglement
  growth after inhomogenous quenches}, Phys. Rev. B 100 (2019) 125139.
\newblock \href {http://dx.doi.org/10.1103/PhysRevB.100.125139}
  {\path{doi:10.1103/PhysRevB.100.125139}}.
\newline\urlprefix\url{https://link.aps.org/doi/10.1103/PhysRevB.100.125139}

\bibitem{parez2020}
G.~Parez, R.~Bonsignori, P.~Calabrese, Quasiparticle dynamics of symmetry
  resolved entanglement after a quench: the examples of conformal field
  theories and free fermions\href {http://arxiv.org/abs/2010.09794}
  {\path{arXiv:2010.09794}}.

\bibitem{SongRachel}
H.~F. Song, S.~Rachel, C.~Flindt, I.~Klich, N.~Laflorencie, K.~Le~Hur,
  \href{https://link.aps.org/doi/10.1103/PhysRevB.85.035409}{Bipartite
  fluctuations as a probe of many-body entanglement}, Phys. Rev. B 85 (2012)
  035409.
\newblock \href {http://dx.doi.org/10.1103/PhysRevB.85.035409}
  {\path{doi:10.1103/PhysRevB.85.035409}}.
\newline\urlprefix\url{https://link.aps.org/doi/10.1103/PhysRevB.85.035409}

\bibitem{Bonsignori2019}
R.~Bonsignori, P.~Ruggiero, P.~Calabrese,
  \href{https://doi.org/10.1088%2F1751-8121%2Fab4b77}{Symmetry resolved
  entanglement in free fermionic systems}, Journal of Physics A: Mathematical
  and Theoretical 52~(47) (2019) 475302.
\newblock \href {http://dx.doi.org/10.1088/1751-8121/ab4b77}
  {\path{doi:10.1088/1751-8121/ab4b77}}.
\newline\urlprefix\url{https://doi.org/10.1088%2F1751-8121%2Fab4b77}

\bibitem{MurcianodiGiulio}
S.~Murciano, G.~D. Giulio, P.~Calabrese,
  \href{https://scipost.org/10.21468/SciPostPhys.8.3.046}{{Symmetry resolved
  entanglement in gapped integrable systems: a corner transfer matrix
  approach}}, SciPost Phys. 8 (2020) 46.
\newblock \href {http://dx.doi.org/10.21468/SciPostPhys.8.3.046}
  {\path{doi:10.21468/SciPostPhys.8.3.046}}.
\newline\urlprefix\url{https://scipost.org/10.21468/SciPostPhys.8.3.046}

\bibitem{MurcianodiGiulio2}
S.~Murciano, G.~D. Giulio, P.~Calabrese, Entanglement and symmetry resolution
  in two dimensional free quantum field theories, JHEP 2020 (2020) 73.

\bibitem{LukinRispoli}
A.~Lukin, M.~Rispoli, R.~Schittko, M.~E. Tai, A.~M. Kaufman, S.~Choi,
  V.~Khemani, J.~Leonard, M.~Greiner, Probing entanglement in a
  many-body-localized system, Science 364 (2019) 256.

\bibitem{BrydgesElben}
T.~Bridges, A.~Elben, P.~Jurcevic, B.~Vermersch, C.~Maier, B.~P. Lanyon,
  P.~Zoller, R.~Blatt, C.~F. Roos, Probing rényi entanglement entropy via
  randomized measurements, Science 364 (2019) 260.
\newblock \href {http://dx.doi.org/10.1126/science.aau4963}
  {\path{doi:10.1126/science.aau4963}}.

\bibitem{Trotter}
H.~F. Trotter, On the product of semi-groups of operators, Proc. Amer. Math.
  Soc. 10 (1959) 545.

\bibitem{Suzuki1}
M.~Suzuki, Generalized trotter's formula and systematic approximants of
  exponential operators and inner derivations with applications to many-body
  problems, Commun. Math. Phys. 51 (1976) 183.

\bibitem{Suzuki2}
M.~Suzuki, Transfer-matrix method and monte carlo simulation in quantum spin
  systems, Phys. Rev. B 31 (1985) 2957.

\bibitem{Luitz2}
D.~J. Luitz, N.~Laflorencie, F.~Alet, Extended slow dynamical regime close to
  the many-body localization transition, Phys. Rev. B 93 (2016) 060201.

\bibitem{Gopalakrishnan2015}
S.~Gopalakrishnan, M.~M\"uller, V.~Khemani, M.~Knap, E.~Demler, D.~A. Huse,
  \href{https://link.aps.org/doi/10.1103/PhysRevB.92.104202}{Low-frequency
  conductivity in many-body localized systems}, Phys. Rev. B 92 (2015) 104202.
\newblock \href {http://dx.doi.org/10.1103/PhysRevB.92.104202}
  {\path{doi:10.1103/PhysRevB.92.104202}}.
\newline\urlprefix\url{https://link.aps.org/doi/10.1103/PhysRevB.92.104202}

\bibitem{Agarwal2017}
K.~Agarwal, E.~Altman, E.~Demler, S.~Gopalakrishnan, D.~A. Huse, M.~Knap,
  \href{https://onlinelibrary.wiley.com/doi/abs/10.1002/andp.201600326}{Rare-region
  effects and dynamics near the many-body localization transition}, Annalen der
  Physik 529~(7) (2017) 1600326.
\newblock \href
  {http://arxiv.org/abs/https://onlinelibrary.wiley.com/doi/pdf/10.1002/andp.201600326}
  {\path{arXiv:https://onlinelibrary.wiley.com/doi/pdf/10.1002/andp.201600326}},
  \href {http://dx.doi.org/https://doi.org/10.1002/andp.201600326}
  {\path{doi:https://doi.org/10.1002/andp.201600326}}.
\newline\urlprefix\url{https://onlinelibrary.wiley.com/doi/abs/10.1002/andp.201600326}

\bibitem{Peschel2004}
I.~Peschel, On the reduced density matrix for a chain of free electrons, J.
  Stat. Mech. (2004) P06004.

\bibitem{PeschelEisler}
I.~Peschel, V.~Eisler, Reduced density matrices and entanglement entropy in
  free lattice models, J. Phys. A 42~(50) (2009) 504003.

\bibitem{KlichLevitov09}
I.~Klich, L.~Levitov,
  \href{https://link.aps.org/doi/10.1103/PhysRevLett.102.100502}{Quantum noise
  as an entanglement meter}, Phys. Rev. Lett. 102 (2009) 100502.
\newblock \href {http://dx.doi.org/10.1103/PhysRevLett.102.100502}
  {\path{doi:10.1103/PhysRevLett.102.100502}}.
\newline\urlprefix\url{https://link.aps.org/doi/10.1103/PhysRevLett.102.100502}

\bibitem{CalabreseMintchev}
P.~Calabrese, M.~Mintchev, E.~Vicari,
  \href{https://doi.org/10.1209%2F0295-5075%2F98%2F20003}{Exact relations
  between particle fluctuations and entanglement in fermi gases}, {EPL}
  (Europhysics Letters) 98~(2) (2012) 20003.
\newblock \href {http://dx.doi.org/10.1209/0295-5075/98/20003}
  {\path{doi:10.1209/0295-5075/98/20003}}.
\newline\urlprefix\url{https://doi.org/10.1209%2F0295-5075%2F98%2F20003}

\bibitem{Cover1991}
T.~M. Cover, J.~A. Thomas, Elements of information theory, Wiley, New York,
  (1991).

\bibitem{Klich2006}
I.~Klich, \href{https://doi.org/10.1088%2F0305-4470%2F39%2F4%2Fl02}{Lower
  entropy bounds and particle number fluctuations in a fermi sea}, Journal of
  Physics A: Mathematical and General 39~(4) (2006) L85--L91.
\newblock \href {http://dx.doi.org/10.1088/0305-4470/39/4/l02}
  {\path{doi:10.1088/0305-4470/39/4/l02}}.
\newline\urlprefix\url{https://doi.org/10.1088%2F0305-4470%2F39%2F4%2Fl02}

\bibitem{Muth2011}
D.~Muth, R.~G. Unanyan, M.~Fleischhauer,
  \href{https://link.aps.org/doi/10.1103/PhysRevLett.106.077202}{Dynamical
  simulation of integrable and nonintegrable models in the heisenberg picture},
  Phys. Rev. Lett. 106 (2011) 077202.
\newblock \href {http://dx.doi.org/10.1103/PhysRevLett.106.077202}
  {\path{doi:10.1103/PhysRevLett.106.077202}}.
\newline\urlprefix\url{https://link.aps.org/doi/10.1103/PhysRevLett.106.077202}

\bibitem{ZhaoAndraschkoSirker}
Y.~Zhao, F.~Andraschko, J.~Sirker, Entanglement entropy of disordered quantum
  chains following a global quench, Phys. Rev. B 93 (2016) 205146.

\bibitem{ZhaoSirker2020}
Y.~Zhao, D.~Feng, Y.~Hu, S.~Guo, J.~Sirker,
  \href{https://link.aps.org/doi/10.1103/PhysRevB.102.195132}{Entanglement
  dynamics in the three-dimensional anderson model}, Phys. Rev. B 102 (2020)
  195132.
\newblock \href {http://dx.doi.org/10.1103/PhysRevB.102.195132}
  {\path{doi:10.1103/PhysRevB.102.195132}}.
\newline\urlprefix\url{https://link.aps.org/doi/10.1103/PhysRevB.102.195132}

\bibitem{ChoiHild}
J.-y. Choi, S.~Hild, J.~Zeiher, P.~Schau{\ss}, A.~Rubio-Abadal, T.~Yefsah,
  V.~Khemani, D.~A. Huse, I.~Bloch, C.~Gross,
  \href{https://science.sciencemag.org/content/352/6293/1547}{Exploring the
  many-body localization transition in two dimensions}, Science 352~(6293)
  (2016) 1547--1552.
\newblock \href
  {http://arxiv.org/abs/https://science.sciencemag.org/content/352/6293/1547.full.pdf}
  {\path{arXiv:https://science.sciencemag.org/content/352/6293/1547.full.pdf}},
  \href {http://dx.doi.org/10.1126/science.aaf8834}
  {\path{doi:10.1126/science.aaf8834}}.
\newline\urlprefix\url{https://science.sciencemag.org/content/352/6293/1547}

\bibitem{WahlPal}
T.~B. Wahl, A.~Pal, S.~H. Simon, Signatures of the many-body localized regime
  in two dimensions, Nat. Phys. 15 (2019) 164.

\bibitem{Grozdanov2015}
S.~c.~v. Grozdanov, A.~Lucas, S.~Sachdev, K.~Schalm,
  \href{https://link.aps.org/doi/10.1103/PhysRevLett.115.221601}{Absence of
  disorder-driven metal-insulator transitions in simple holographic models},
  Phys. Rev. Lett. 115 (2015) 221601.
\newblock \href {http://dx.doi.org/10.1103/PhysRevLett.115.221601}
  {\path{doi:10.1103/PhysRevLett.115.221601}}.
\newline\urlprefix\url{https://link.aps.org/doi/10.1103/PhysRevLett.115.221601}

\bibitem{Grozdanov2016}
S.~c.~v. Grozdanov, A.~Lucas, K.~Schalm,
  \href{https://link.aps.org/doi/10.1103/PhysRevD.93.061901}{Incoherent thermal
  transport from dirty black holes}, Phys. Rev. D 93 (2016) 061901.
\newblock \href {http://dx.doi.org/10.1103/PhysRevD.93.061901}
  {\path{doi:10.1103/PhysRevD.93.061901}}.
\newline\urlprefix\url{https://link.aps.org/doi/10.1103/PhysRevD.93.061901}

\end{thebibliography}

\end{document}